\documentclass[useAMS,usenatbib]{mn2e}
\usepackage[a4paper,margin=2cm]{geometry}
\usepackage{times}
\usepackage[T1]{fontenc}
\usepackage{aecompl}
\usepackage{epsfig}
\usepackage[]{graphicx} 
\usepackage{pdfpages}
\usepackage{verbatim} 
\usepackage[fleqn]{amsmath}
\usepackage{amssymb}
\usepackage{calc}
\usepackage{caption}
\usepackage[normal]{threeparttable}
\usepackage{longtable}
\usepackage{pdflscape}
\newcommand{\comments}[1]{} 
\usepackage{placeins} 
\usepackage{float}
\usepackage{soul}
\graphicspath{{./plots/}}
\title[HI gas in a young radio galaxy using ASKAP]{Discovery of
  \mbox{H\,{\sc i}} gas in a young radio galaxy at $\bmath{z = 0.44}$
  using the Australian Square Kilometre Array Pathfinder}
\author[J.~R. Allison et al.]{J.~R. Allison$^{1}$\thanks{E-mail:
    james.allison@csiro.au}, E.~M. Sadler$^{2,3}$, V.~A. Moss$^{2,3}$,
  M.~T. Whiting$^{1}$, R.~W. Hunstead$^{2}$, \newauthor
  M.~B. Pracy$^{2}$, S.~J. Curran$^{2,3,4}$, S.~M. Croom$^{2,3}$,
  M. Glowacki$^{1,2,3}$
% ...
\\\\Author
  affiliations given at the end of the paper}
\begin{document}

\date{}

\pagerange{\pageref{firstpage}--\pageref{lastpage}} \pubyear{2015}

\maketitle

\label{firstpage}

\begin{abstract} We report the discovery of a new 21-cm \mbox{H\,{\sc
      i}} absorption system using commissioning data from the Boolardy
  Engineering Test Array of the Australian Square Kilometre Array
  Pathfinder (ASKAP). Using the 711.5 -- 1015.5\,MHz band of ASKAP we
  were able to conduct a blind search for the 21-cm line in a
  continuous redshift range between $z = 0.4$ and $1.0$, which has,
  until now, remained largely unexplored. The absorption line is
  detected at $z = 0.44$ towards the GHz-peaked spectrum radio source
  PKS\,B1740$-$517 and demonstrates ASKAP's excellent capability for
  performing a future wide-field survey for \mbox{H\,{\sc i}}
  absorption at these redshifts. Optical spectroscopy and imaging
  using the Gemini-South telescope indicates that the \mbox{H\,{\sc
      i}} gas is intrinsic to the host galaxy of the radio source.
  The narrow \mbox{[O\,{\sc iii}]} emission lines show clear
  double-peaked structure, indicating either large-scale outflow or
  rotation of the ionized gas. Archival data from the
  \emph{XMM-Newton} satellite exhibit an absorbed X-ray spectrum that
  is consistent with a high column density obscuring medium around the
  active galactic nucleus. The \mbox{H\,{\sc i}} absorption profile is
  complex, with four distinct components ranging in width from 5 to
  300\,km\,s$^{-1}$ and fractional depths from 0.2 to
  20\,per\,cent. In addition to systemic \mbox{H\,{\sc i}} gas, in a
  circumnuclear disc or ring structure aligned with the radio jet, we
  find evidence for a possible broad outflow of neutral gas moving at
  a radial velocity of $v \sim 300$\,km\,s$^{-1}$. We infer that the
  expanding young radio source ($t_{\rm age} \approx 2500$\,yr) is
  cocooned within a dense medium and may be driving circumnuclear
  neutral gas in an outflow of
  $\sim$\,1\,$\mathrm{M}_{\odot}$\,yr$^{-1}$.
  \end{abstract}

  \begin{keywords} {methods: data analysis -- ISM: jets and outflows
      -- galaxies: active -- galaxies: ISM -- radio lines: galaxies.}
\end{keywords}

\section{Introduction}\label{section:introduction}

Over the past 60\,yr the 21-cm hyperfine transition of atomic hydrogen
(\mbox{H\,{\sc i}}) has been used by astronomers to measure the
distribution and kinematics of the neutral interstellar medium (ISM)
in galaxies and, importantly, trace the available fuel for future
star-formation. Wide-field surveys have systematically measured the
\mbox{H\,{\sc i}} content in nearby galaxies
(e.g. \citealt{VanDerHulst:2001, Meyer:2004, Walter:2008,
  Koribalski:2010, Oosterloo:2010b, Haynes:2011, Heald:2011,
  Serra:2012, Wang:2013}), while at greater distances individual
\mbox{H\,{\sc i}}-rich galaxies have been detected out to $z \sim 0.3$
(e.g. \citealt{Catinella:2008,Verheijen:2010,Freudling:2011}).
Statistical techniques, such as spectral stacking
(e.g. \citealt{Lah:2009, Delhaize:2013, Rhee:2013, Gereb:2014}) and
intensity mapping (e.g. \citealt{Chang:2010, Masui:2013}) have also
provided constraints on the cosmological \mbox{H\,{\sc i}} mass
density at $z < 0.4$ and $z < 1.0$ respectively.

At higher redshifts it becomes increasingly difficult for existing
radio telescopes to detect the faint 21-cm emission from individual
galaxies and we must look to absorption against background sources,
either at 21-cm (e.g. \citealt{Carilli:1998, Curran:2006a,
  Curran:2008, Curran:2011a, Curran:2013a, Kanekar:2009a}) or the
Ly$\alpha$ line at 1216\,\AA~(e.g. \citealt{Peroux:2003, Rao:2006,
  Noterdaeme:2012, Zafar:2013}), to understand the evolving
\mbox{H\,{\sc i}} content of the distant Universe. Deficiency of
spatial information is the significant drawback of using absorption to
map the neutral ISM, yet it still provides one of the few practical
methods of directly detecting cold \mbox{H\,{\sc i}} gas in individual
galaxies at an epoch of the Universe ($0.5 \lesssim z \lesssim 2$)
that has, until now, remained largely unexplored
(e.g. \citealt{Lagos:2014}).

Aside from intervening galaxies, \mbox{H\,{\sc i}} absorption is also
a direct tracer of the interaction between radio source and host
galaxy. Recent 21-cm surveys of radio-loud active galactic nuclei
(AGN) have revealed the varied kinematic signatures of rotating discs,
infalling and outflowing gas, and individual offset clouds
(e.g. \citealt{VanGorkom:1989, Morganti:2001, Vermeulen:2003,
  Morganti:2005b, Gupta:2006a, Emonts:2010, Chandola:2011,
  Allison:2012a, Allison:2014, Gereb:2015}). Amongst these
heterogeneous samples we often find that the most compact sources have
the highest rate of detection, the result of either intrinsically
higher column densities of neutral gas (\citealt{Pihlstrom:2003}), the
relative orientation between source and absorber
(\citealt{Curran:2013b}) or, quite possibly, a combination of both
(\citealt{Orienti:2006, Gereb:2015}).  The population of compact
steep-spectrum (CSS; $d \lesssim 15$\,kpc) and gigahertz
peaked-spectrum (GPS; $d \lesssim 1$\,kpc) radio sources are
particularly interesting targets since they are thought to constitute
recently triggered radio AGN ($t_{\rm age} \sim 10^{4}$ --
$10^{6}$\,yr;
\citealt{Fanti:1995,Readhead:1996,Owsianik:1998}). Higher detection
yields of \mbox{H\,{\sc i}} absorption in these sources may be
indicative of the dense and dusty environments in which they are born.

By studying the kinematics of the \mbox{H\,{\sc i}} gas in young radio
galaxies, along with the molecular and ionized gas, we can understand
the processes by which these radio jets are triggered and their
subsequent feedback on the neutral ISM. \cite{Vermeulen:2003} and
\cite{Morganti:2005b} found examples of blue-shifted \mbox{H\,{\sc i}}
absorption components at large negative velocities ($\sim
-1000$\,km\,s$^{-1}$) with respect to galaxies hosting compact radio
sources, symptomatic of jet-driven outflows of neutral gas. More
recently, in-depth case studies of individual radio AGN, using high
spatial resolution 21-cm and multiwavelength data, have provided
direct evidence for the existence of $\sim
10\,\mathrm{M}_{\odot}$\,yr$^{-1}$ jet-driven outflows of
\mbox{H\,{\sc i}}, co-located with outflows of ionized and molecular
gas e.g. 1504+377 (\citealt{Kanekar:2008}), 4C\,12.50
(\citealt{Morganti:2013}), 3C\,293 (\citealt{Mahony:2013}) and
IC\,5063 (\citealt{Tadhunter:2014}).

The advent of precursor and pathfinder telescopes to the Square
Kilometre Array (SKA), in particular the Australian SKA Pathfinder
(ASKAP; \citealt{Johnston:2007,Deboer:2009,Schinckel:2012}), the South
African MeerKAT telescope (\citealt{Booth:2009}) and the Westerbork
Aperture Tile in Focus (APERTIF; \citealt{Oosterloo:2009}), will
enable astronomers to carry out radio-selected surveys for
\mbox{H\,{\sc i}} absorption over most of the sky. Crucially, such
surveys are made possible by significant improvements in bandwidth,
field-of-view, and new observatories with reduced terrestrial radio
frequency interference (RFI).

Here we report the first result from a search for redshifted
absorption using the ASKAP Boolardy Engineering Test Array (BETA;
\citealt{Hotan:2014}), a six-antenna prototype designed to demonstrate
the feasibility and science capability of phased-array feeds (PAFs)
based on a planar connected ``chequerboard'' array
(\citealt{Hay:2008}). The backend digital beamformer weights the
signal from 188 independent PAF receptors at the focal plane of each
antenna to electronically form up to nine simultaneous primary beams
within an area of approximately 30\,deg$^{2}$. This impressive
flexibility in field-of-view is matched by the telescope's spectral
capability; the fine filterbank generates 16\,416 channels over
304\,MHz of bandwidth, at observing frequencies between 0.7 and
1.8\,GHz, achieving an average 21-cm line resolution of
$5.5$\,km\,s$^{-1}$ and maximum \mbox{H\,{\sc i}} redshift of $z =
1.0$. We are currently using BETA to search for \mbox{H\,{\sc i}}
absorption towards the brightest and most compact radio sources in the
southern sky. We will continue to modify our target sample to fainter
sources as more antennas are added to the ASKAP array, culminating in
the First Large Absorption Survey in \mbox{H\,{\sc i}} (FLASH).

Throughout this paper we adopt a flat $\Lambda$ cold dark matter
($\Lambda$CDM) cosmology with $H_{0}$ = 70\,km\,s$^{-1}$\,Mpc$^{-1}$,
$\Omega_\mathrm{M}$ = 0.3 and $\Omega_{\Lambda}$ = 0.7. Radial
velocities and redshifts have been corrected for the solar barycentric
standard of rest frame. Uncertainties are given as 1\,$\sigma$
intervals unless otherwise stated.

\section{HI absorption with BETA}

\subsection{A pilot survey of bright, compact radio sources}

As part of the early commissioning and science demonstration phase of
ASKAP, we are using BETA to carry out a pilot survey of \mbox{H\,{\sc
    i}} absorption at $0.4 < z < 1.0$ towards the brightest and most
compact radio sources in the southern sky.  The sensitivity of the
telescope is substantially lower than will be achieved with the full
ASKAP\footnote{The full array is expected to have 36 antennas, each
  fitted with a PAF that can simultaneously form up to 36 beams.}  and
so we select targets that optimize the optical depth sensitivity,
through both a high signal-to-noise ratio (S/N) continuum background
and high expected fraction of radio flux obscured by foreground
neutral gas. The following is a list of criteria from which we select
our targets.
\begin{enumerate}
\item Visible at declinations of $\delta < +20\degr$.
\item A total flux density greater than 1\,Jy in existing all-sky
  catalogues at similar frequencies to this band, which include the
  National Radio Astronomy Observatory Very Large Array Sky Survey
  (NVSS; \citealt{Condon:1998}), the Sydney University Molonglo Sky
  Survey (SUMSS; \citealt{Mauch:2003}) and the second epoch Molonglo
  Galactic Plane Survey (MGPS-2; \citealt{Murphy:2007}).
\item Either $z \geq 0.4$ or no known redshift according to the
  NASA/IPAC Extragalactic Database
  (NED)\footnote{\url{http://ned.ipac.caltech.edu/}}, thus excluding
  those sources that are known to be located in front of the volume
  probed by $0.4 < z < 1.0$.
\item A significant fraction of the total flux density is distributed
  on the compact scales measured by very-long-baseline-interferometry
  (VLBI), based on those sources listed in the VLBI radio fundamental
  catalogue\footnote{\url{http://astrogeo.org/rfc} (\citealt{
      Petrov:2013} and references therein).}.
\end{enumerate}
We present here our first new detection of 21-cm absorption using BETA
as a demonstration of the capability of the ASKAP telescope to perform
radio-selected surveys for redshifted \mbox{H\,{\sc i}}. In future
work we will present the results of further observations from this
pilot survey.

\begin{figure}
\centering
\includegraphics[width = 0.5\textwidth]{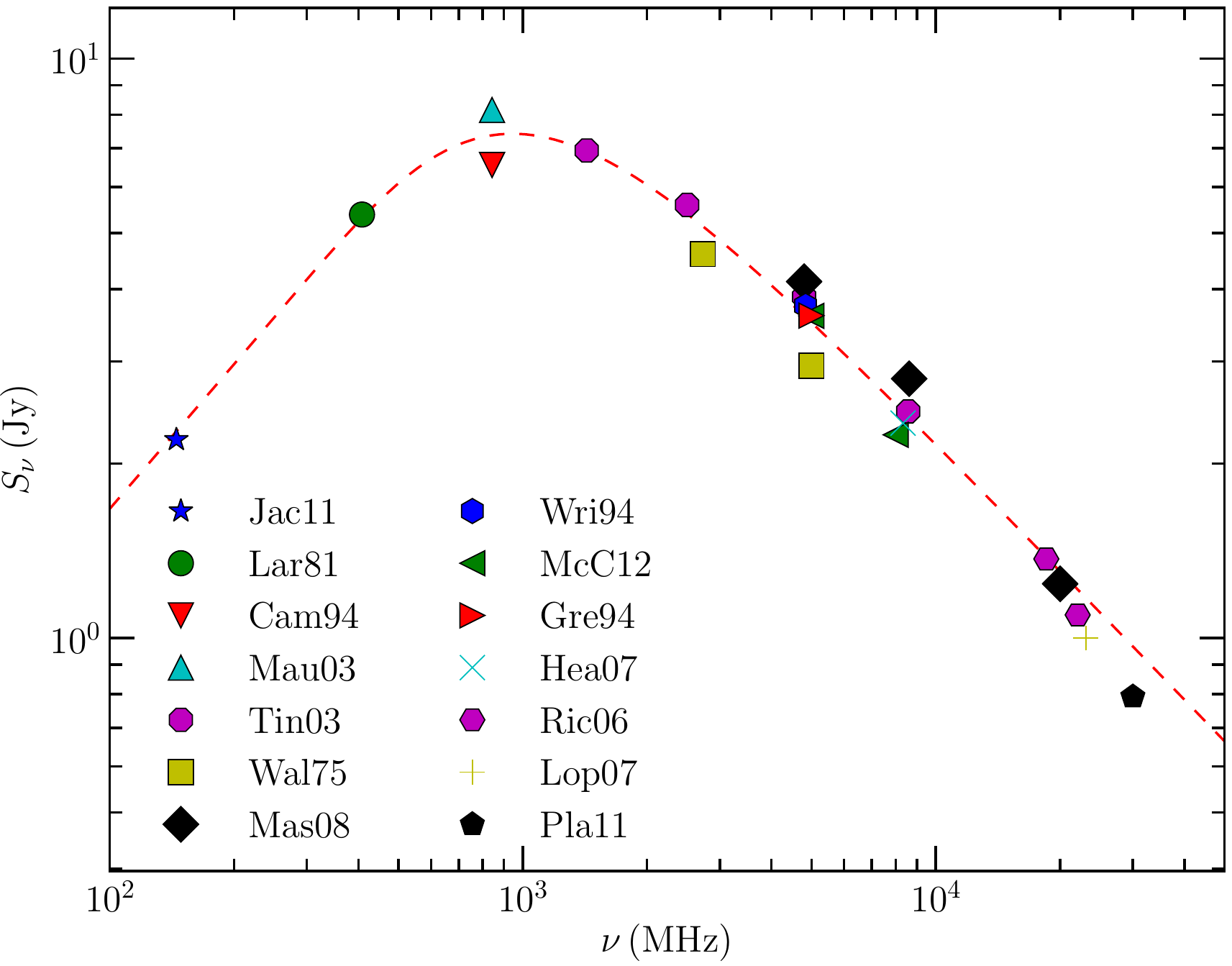}
\caption{The SED at radio wavelengths of GPS radio source
  PKS\,B1740$-$517. The frequency axis is given in the observer
  rest-frame. The dashed line denotes a best-fitting broken power-law
  model to the data (\citealt{Moffet:1975}). Using standard least
  squares minimization we obtain a break frequency of 1\,GHz and
  optically thick and thin spectral indices of $\alpha = 0.83$ and
  $-0.74$, respectively. References for the data: Jac11 --
  \citet{Jacobs:2011}; Lar81 -- \citet{Large:1981}; Cam94 --
  \citet{Campbell-Wilson:1994}; Mau03 -- \citet{Mauch:2003}; Tin03 --
  \citet{Tingay:2003}; Wal75 -- \citet{Wall:1975}; Mas08 --
  \citet{Massardi:2008}; Wri94 -- \citet{Wright:1994}; McC12 --
  \citet{McConnell:2012}; \citet{Gregory:1994} -- Gre94; Hea07 --
  \citet{Healey:2007}; Ric06 -- \citet{Ricci:2006}; Lop07 --
  \citet{Lopez-Caniego:2007}; Pla11 --
  \citet{Planck:2011}.}\label{figure:PKS1740-517_radio_sed}
\end{figure}

\begin{figure*}
\centering
\includegraphics[width = 0.525\textwidth]{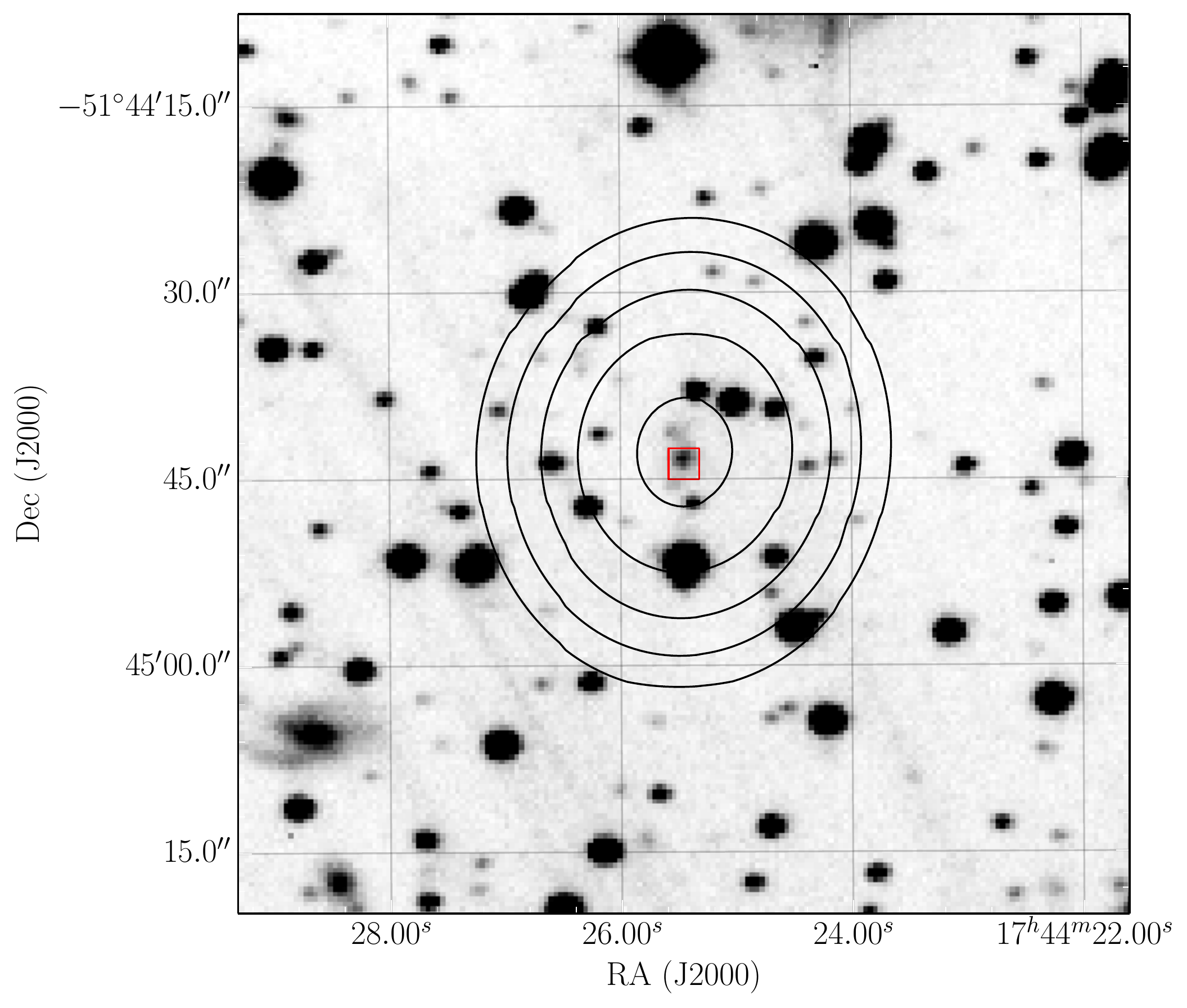}
\quad
\includegraphics[width = 0.445\textwidth]{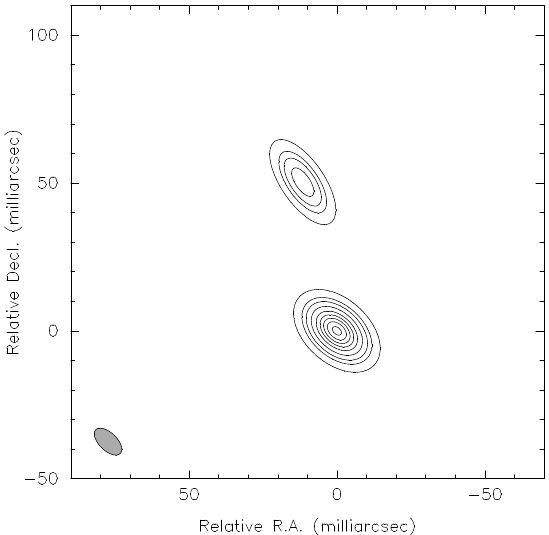}
\caption{Left: an $R$-band optical image centred on PKS\,B1740$-$517
  (300\,s exposure with the 3.9\,m Anglo-Australian Telescope;
  \citealt{Burgess:2006}), overlaid with SUMSS 843\,MHz radio contours
  (5.5, 6.0, 6.5, 7.0 and 7.5\,Jy\,beam$^{-1}$;
  \citealt{Mauch:2003}). The open red square denotes the position of
  the radio source -- $\rmn{RA}(\rmn{J}2000) = 17 ^{\rmn{h}}
  44^{\rmn{m}} 25\fs450$ and $\rmn{Dec.}(\rmn{J}2000) = -51\degr
  44\arcmin 43\farcs79$ -- given by the second realization of the
  International Celestial Reference Frame (ICRF2;
  \citealt{Fey:2009}). The faint background streaking in the optical
  image is due to scattered light from $\mu$\,Arae, a fifth magnitude
  star at an angular separation of 5.9\,arcmin to the
  south-west. Right: a high spatial resolution 2.3\,GHz source model
  of PKS\,B1740$-$517, obtained by \citet{King:1994} from VLBI
  observations using three antenna baselines, taken on 1989 December
  06. The positions are given relative to the centroid of the brighter
  component. The contour levels are at 1, 5, 10, 20, 35, 50, 65, 80
  and 95 per\,cent of the 2.67\,Jy\,beam$^{-1}$ peak flux density. The
  model components have been convolved by the synthesized beam given
  by the shaded ellipse in the bottom left-hand corner. The source is
  well modelled by two hotspot components separated by 52\,mas in a
  flux ratio of 3.9 to 1.}\label{figure:PKS1740-517_images}
\end{figure*}

\subsection{PKS\,B1740$-$517}\label{section:pks1740-517}

PKS\,B1740$-$517 was one of the first sources selected for BETA
observations based on the selection criteria outlined above. It is an
archetypal GPS radio source (\citealt{Randall:2011}), peaking at an
observed frequency of 1\,GHz, with optically thick and thin spectral
indices\footnote{We adopt the convention that the spectral index
  $\alpha$ is defined by $S \propto \nu^{\alpha}$, where $S$ is the
  flux density and $\nu$ is the frequency.} of $\alpha = 0.83$ and
$-0.74$, respectively (see
Fig.\,\ref{figure:PKS1740-517_radio_sed}). The relative compactness
and high flux density ($S_{843} = 8.15 \pm 0.24$\,Jy;
\citealt{Mauch:2003}) of this source make it an ideal target for
\mbox{H\,{\sc i}} absorption. Flux monitoring by King et al. over
$\sim$700\,d, using the 26-m antenna at the Mt. Pleasant Observatory
(\citealt{King:1994}; \citealt{Jauncey:2003}), indicated that the
source is weakly varying at 2.3\,GHz (by approximately 20\,per\,cent),
with no variation seen at 8.4\,GHz over this time range. They
demonstrated, by fitting to visibility data from three baselines in
the network of Australian VLBI antennas (Parkes -- Hobart, Parkes --
Perth and Parkes -- Alice Springs), that the 2.3\,GHz continuum
emission is adequately modelled by two compact Gaussian components of
angular extent less than 10\,mas and separated by 52\,mas
(Fig.\,\ref{figure:PKS1740-517_images}). While there is some residual
flux on the longest baselines, which might be indicative of more
complex compact structure, the total flux density at 2.3\,GHz is
apparently accounted for by the modelled compact components. The
spectral behaviour, VLBI-scale structure and low variability are all
evidence that this radio source is intrinsically young and/or confined
by its environment (e.g. \citealt{Odea:1998}).

At a Galactic latitude of $b = -11\fdg5$ and longitude of $l =
340\fdg2$, PKS\,B1740$-$517 is seen through a densely populated
foreground (Fig.\,\ref{figure:PKS1740-517_images}), which is likely
the reason for the lack of optical information for this
source. Despite this, \cite{DiSerego-Alighieri:1994} used imaging and
spectroscopy with the ESO 3.6-m telescope to secure an optical
identification of the host galaxy, noting that it has a particularly
red continuum. Unfortunately they could not detect any strong optical
emission lines and hence secure a spectroscopic redshift. Approximate
indicators of the redshift have been determined by photometric means:
\cite{Wall:1985} give a rough estimate of $z = 0.347$ based on their
empirically derived $V$-band magnitude relationship (but noting a poor
optical identification with the radio source), while
\cite{Burgess:2006} estimate that $z = 0.63$ based on the $R$-band
magnitude of 20.8. At longer wavelengths, the infrared colours [3.4 --
4.6 \,$\mu$\,m] = 1.06 and [4.6 -- 12\,$\mu$\,m] = 2.85, from the
\emph{Wide-field Infrared Survey Explorer} (\emph{WISE};
\citealt{Wright:2010}), are consistent with a quasi-stellar object at
$z \sim 0.5$ (fig.\,1 of \citealt{Blain:2013}). These photometric
indicators of the redshift suggested that PKS\,B1740$-$517 would
provide a suitable background target against which we might detect
absorption in the range $0.4 < z < 1.0$ with BETA.

\subsection{Observations}

\begin{figure*}
\centering
\includegraphics[width = 1.0\textwidth]{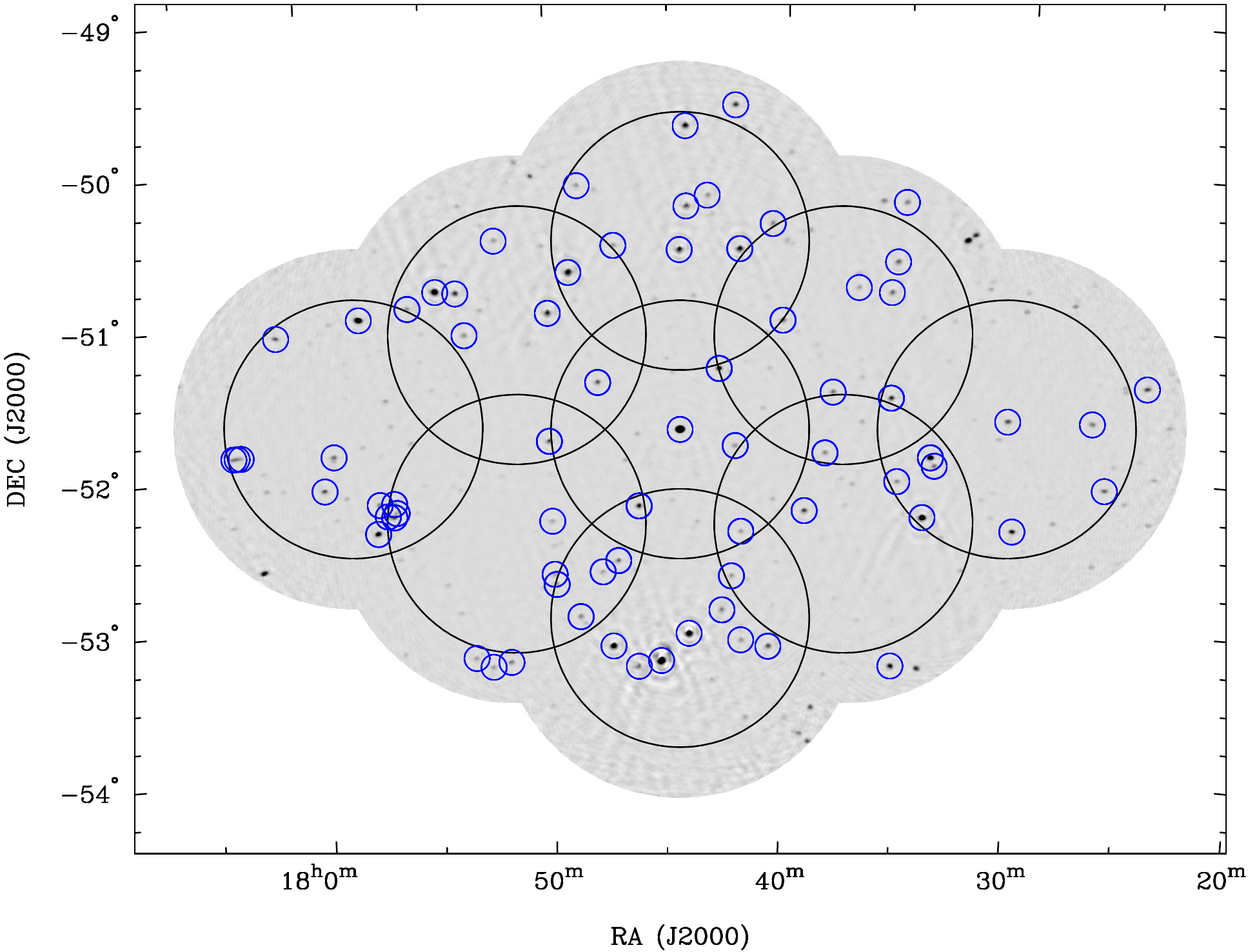}
\caption{A total intensity image of the PKS\,B1740$-$517 field
  observed with BETA (grey-scale; data obtained from a commissioning
  observation carried out on 2014 September 01). The small blue
  circles denote the positions of sources brighter than 100\,mJy
  within 1\degr of each beam centre, listed in the 843\,MHz SUMSS
  (\citealt{Mauch:2003}) and MGPS-2 (\citealt{Murphy:2007})
  catalogues. The large black circles denote the nominal beam width at
  half power at 863.5\,MHz ($1.028\lambda/D_\mathrm{dish} \approx
  1.7\degr$) for each of the nine beams.}
\label{figure:PKS1740-517_beta_mosaic}
\end{figure*}

\begin{figure*}
\centering
\includegraphics[width = 0.95\textwidth]{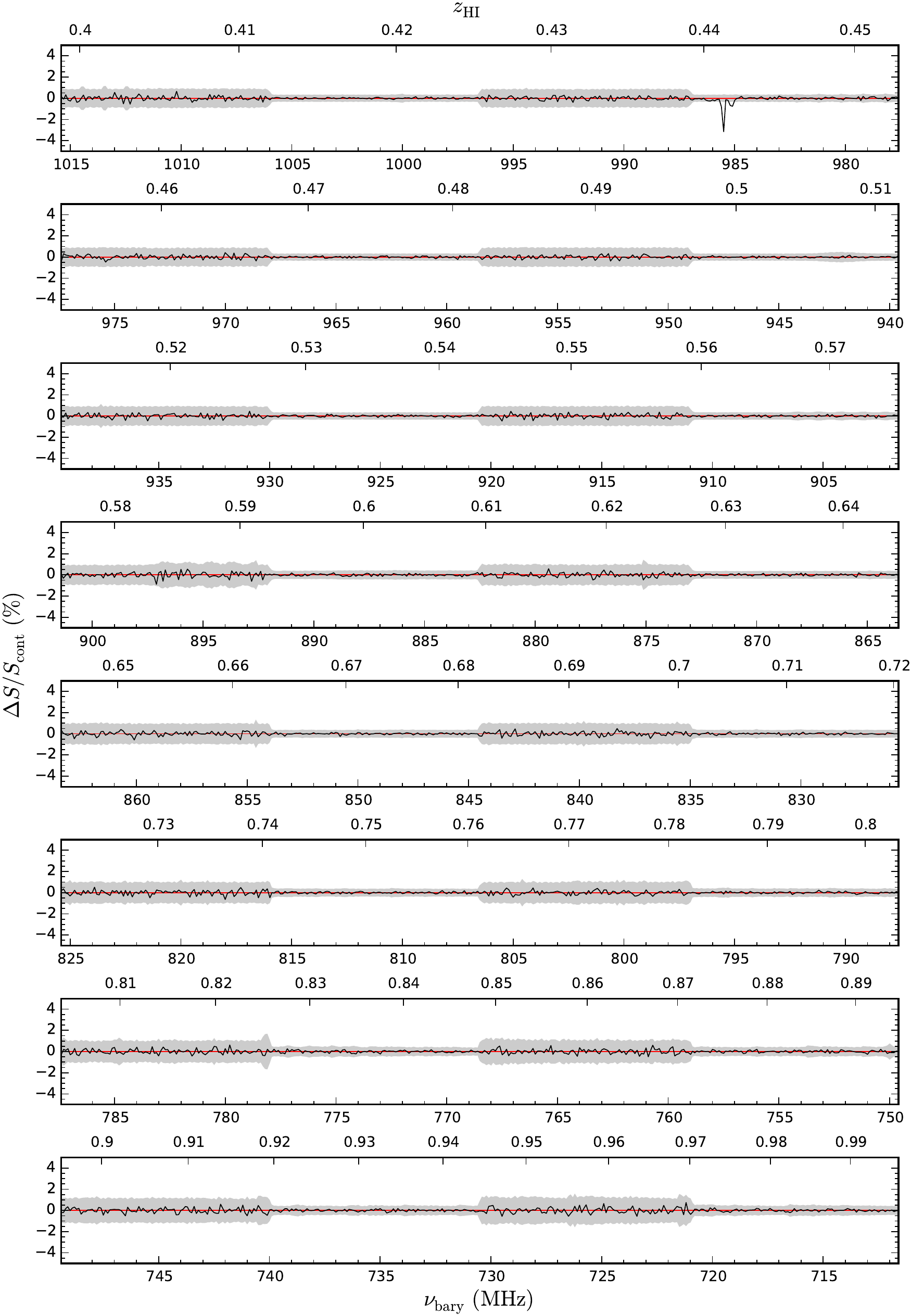}
\caption{The 711.5--1015.5\,MHz BETA spectrum towards
  PKS\,B1740$-$517, averaged over all three observing epochs. For
  visual clarity the data have been binned from the native spectral
  resolution of 18.5 to 100\,kHz. The barycentric corrected observed
  frequency is shown on the lower abscissa, and the upper-abscissa
  denotes the corresponding \mbox{H\,{\sc i}} redshift. The data
  (black line) denote the change in flux density as a fraction of the
  continuum and the grey region gives the corresponding rms spectral
  noise multiplied by a factor of 5. An absorption line is visible in
  the spectrum at $\nu_\mathrm{bary} = 985.5$\,MHz, equal to a
  \mbox{H\,{\sc i}} redshift of $z = 0.4413$. In
  Fig.\,\ref{figure:PKS1740-517_beta_fit} we show this absorption line
  at full spectral resolution. An increase in the noise is seen in
  those parts of the band where 50\,per\,cent of the correlator was
  unavailable for two of our three observations. Other noise features
  are the result of flagging and occasional failures of individual
  correlator cards.}\label{figure:PKS1740-517_beta_spectrum}
\end{figure*}

On three dates -- 2014 June 24, August 03 and September 01 -- we used
the 711.5--1015.5\,MHz band of the ASKAP BETA prototype
(\citealt{Hotan:2014}) to search for 21-cm absorption towards
PKS\,B1740$-$517 in the redshift range $0.4 < z < 1.0$. In
Table\,\ref{table:beta_observations} we summarize the parameters for
each of these observations. Fig.\,\ref{figure:PKS1740-517_beta_mosaic}
shows the nominal footprint of the nine PAF beams for our observation
(a symmetric diamond pattern), centred on $\rmn{RA}(\rmn{J}2000) =
17^{\rmn{h}} 44^{\rmn{m}} 25\fs45$ and $\rmn{Dec.}(\rmn{J}2000) =
-51\degr 44\arcmin 43\farcs8$. Additional observations of a calibrator
source (PKS\,B1934$-$638) were carried out in short 5 -- 15\,min scans
at the centre of each PAF beam. Using weights for the PAF elements
that maximize the S/N in the required direction (see
\citealt{Hotan:2014}), beams were generated with nominal half-power
widths that decrease from 2\fdg07 to 1\fdg45 across the whole
bandpass. The six 12\,m antennas of BETA are arranged in an array that
is elongated by a factor of approximately 2 in the east$-$west
direction, with baselines in the range 37 -- 916\,m. Therefore the
data are sensitive to angular scales in the range 1 -- 39\,arcmin
across the band. The fine filterbanks in the beamformer generate
16\,416 independent spectral channels separated by approximately
18.5\,kHz, equivalent to \mbox{H\,{\sc i}} velocities in the range
$5.5$ -- $7.8\,\mathrm{km}\,\mathrm{s}^{-1}$.

\subsection{Data analysis}

We used the \texttt{CASA}\footnote{\url{http://casa.nrao.edu}} package
(\citealt{McMullin:2007}) to separate the data into each of the nine
beams and flag the auto-correlations and large amplitude values due to
digital glitches. The PAF beams are formed digitally by applying
weights to elemental receptors, which in the case of BETA are constant
in a repeating pattern of 4 and 5 MHz intervals. There are some sharp
bandpass effects at the edges of these intervals, but within them we
find that the bandpass is smooth for most of the band at a level
consistent with the noise in our data. This is further corroborated by
the absence of any false positive detection; see
Section\,\ref{section:other_sources}. We therefore pursued a
calibration strategy that splits the band, at full spectral
resolution, into 64 chunks based on the exact pattern of 4 and 5\,MHz
intervals. Separately we generated a single data set of coarse $32
\times 9.5\,\mathrm{MHz}$ channels across the band by averaging over
513 of the fine channels. The 18.5-kHz fine channels are used to
search for \mbox{H\,{\sc i}} absorption, while the 9.5\,MHz coarse
channels produce high S/N images that are used for self-calibration.

We performed calibration, imaging and further flagging of the data
using tasks from the
\texttt{MIRIAD}\footnote{\url{http://www.atnf.csiro.au/computing/software/miriad/}}
package \citep{Sault:1995}. Those 18.5-kHz channels that were
corrupted by narrow-band RFI were flagged using a \texttt{MIRIAD}
implementation of the \texttt{Sum}\texttt{Threshold} method developed
by \cite{Offringa:2010}, which typically resulted in data loss of a
few \,per\,cent. The majority of this low-level RFI is caused by
single-channel spikes, commonly referred to as birdies, which are
generated by the cooling system flow regulator valves associated with
the first generation of phased-array feeds. Further corrupted data,
resulting from problems with individual antennas, antenna baselines or
the backend correlator, were also manually removed at this stage. In
particular, there was an intermittent problem with the time
synchronisation between individual correlator cards that resulted in
50\,per\,cent of the spectral band being flagged for the data obtained
on 2014 June 24 and September 01.

The flux density scale and rough gain corrections for each antenna
were calculated using short scans (between 5 and 15\,min) of
PKS\,B1934$-$638 in each PAF beam, based on the model of
\cite{Reynolds:1994}.  These solutions were then transferred to the
corresponding data in the PKS\,B1740$-$517 field. We achieved further
correction for the time-dependent gains by performing multiple
iterations of self-calibration on the $9.5\,\mathrm{MHz}$
coarse-channel data, which span the whole band and thereby provide
optimal S/N. Initially the SUMSS 843\,MHz radio catalogue
\citep{Mauch:2003} was used to construct a sky model for each PAF
beam, followed by models generated from the data using a
multi-frequency synthesis variant of the {\sc CLEAN} algorithm
\citep{Hogbom:1974}. At each iteration the \texttt{MIRIAD} {\sc
  SELFCAL} task was used to solve for the time-dependent gains, and we
simply transferred these corrections to each of the corresponding 64
chunks of 18.5\,kHz fine-channel data.

Precise subtraction of the continuum flux density was achieved by
generating continuum images for each of the 64 chunks of fine-channel
data and then subtracting the {\sc CLEAN} component models (using {\sc
  UVMODEL}) from the visibilities. Removal of any residual flux
density, including curvature of the instrumental bandpass towards low
frequencies, was performed by fitting a second-order polynomial to
each chunk and subtracting using {\sc UVLIN}. Channels containing any
detected absorption features were subsequently excluded in iterative
refinement of the above continuum subtraction procedure. We image the
continuum-subtracted data in the standard way, forming cubes that are
equal to the nominal full width at half-maximum (FWHM) of the beams at
the middle of the band. Spectra were then extracted at the positions
of identified continuum target sources within each beam centre and
converted to units of fractional absorption through dividing by the
continuum.

\begin{table*} 
 \begin{threeparttable}
   \caption{Summary of our 21-cm observations using BETA, where column
     1 gives the date of observation; column 2 and 3 the modified
     Julian date (MJD) start and end times; column 4 the number of PAF
     beams used and their footprint; column 5 the ASKAP antennas used;
     column 6 the on-source integration time; columns 7--9 properties
     of the restoring beam, which include the FWHM of the major and
     minor axes, and their position angle, respectively; column 10 the
     median per-channel noise in the centre beam, across the 304\,MHz
     band. Note that differences in the spectral noise between these
     observations are consistent with the integration times and number
     of antennas.}\label{table:beta_observations}
  \begin{tabular}{lllllrrcrc}
    \hline
    \multicolumn{1}{l}{Date} & \multicolumn{2}{c}{MJD} & \multicolumn{1}{l}{PAF beams\tnote{$a$}} & \multicolumn{1}{l}{Antennas\tnote{$b$}} & \multicolumn{1}{c}{$t_\mathrm{int}$} & \multicolumn{1}{c}{$\theta_{\rm max}$} & \multicolumn{1}{c}{$\theta_{\rm min}$} & \multicolumn{1}{c}{$\phi$} & \multicolumn{1}{c}{$\sigma_{\rm chan}$} \\
    \multicolumn{1}{c}{} & \multicolumn{1}{c}{Start} & \multicolumn{1}{c}{End} & \multicolumn{1}{c}{} & \multicolumn{1}{c}{} & \multicolumn{1}{c}{(h)} & \multicolumn{2}{c}{(arcsec)} & \multicolumn{1}{c}{($\degr$)} & \multicolumn{1}{c}{(mJy\,beam$^{-1}$)} \\   
    \hline
    2014 June 24 & 56832.442 & 56832.921 & 1 (centre) & 1, 8, 9, 15 & 11.5 & 90 & 70 & +70 & 23 \\
    2014 August 03 & 56872.649 & 56872.774 & 9 (diamond) & 1, 3, 6, 8, 15 & 3.0 & 340 & 100 & $-$40 & 37 \\
    2014 September 01 & 56901.224 & 56901.723 & 9 (diamond) & 1, 6, 8, 9, 15 & 11.9 & 100 & 80 & $-$90 & 18 \\
    \hline
   \end{tabular}
   \begin{tablenotes}
   \item[$a$] {See Fig.\,\ref{figure:PKS1740-517_beta_mosaic} for
       details of the 9-beam diamond configuration. The centre
       configuration refers to a single beam at the pointing centre.}
   \item[$b$] {See fig.\,2 of \citet{Hotan:2014} for details of the
      BETA antenna positions.}
   \end{tablenotes}
 \end{threeparttable}
\end{table*}

\begin{table*} 
 \begin{threeparttable}
   \caption{A summary of model parameters derived from fitting the
     \mbox{H\,{\sc i}} absorption line seen towards PKS\,B1740$-$517
     for each epoch and the average spectrum. Column 1 gives the
     observation epoch; column 2 the Gaussian component corresponding
     to that shown in Fig.\,\ref{figure:PKS1740-517_beta_fit}; column
     3 the component redshift; column 4 the component rest-frame FWHM;
     column 5 the peak component depth as a fraction of the continuum
     flux density; column 6 the improvement in Bayes evidence for each
     component relative to the noise-only model (see
     \citealt{Allison:2014} and references therein); column 7 the
     reduced chi-squared statistic for the best-fitting model
     parameters after the component is introduced into the
     model. Intervals of 1\,$\sigma$ are given for the measured
     uncertainties, derived from simultaneously fitting all model
     Gaussian components and taking into account the gain response of
     the spectral channels (see text for details).
   }\label{table:model_parameters}
   \renewcommand*\arraystretch{1.2}
   \begin{tabular}{@{}l@{\hspace{0.1in}}ccc@{\hspace{0.3in}}ccc@{}}
     \hline
     \multicolumn{1}{l}{Epoch} & \multicolumn{1}{c}{ID} & \multicolumn{1}{c}{$z_{\rm bary}$} & \multicolumn{1}{c}{$\Delta{v_{50}}$} &  \multicolumn{1}{c}{$(\Delta{S}/S_{\rm cont})_{\rm peak}$} & \multicolumn{1}{c}{$\Delta{\ln(\mathcal{Z})}$} & \multicolumn{1}{c}{$\chi^{2}/\mathrm{d.o.f.}$} \\
     \multicolumn{1}{c}{} & \multicolumn{1}{c}{} & \multicolumn{1}{c}{} & \multicolumn{1}{c}{(km\,s$^{-1}$)} & \multicolumn{1}{c}{(per cent)} & \multicolumn{1}{c}{} & \multicolumn{1}{c}{} \\
     \hline
     2014 June 24 & 1 & $0.44129264_{-0.00000060}^{+0.00000058}$ & $5.15_{-0.21}^{+0.20}$ & $-20.20_{-0.74}^{+0.68}$ & $893.47\pm 0.08$ & $3.29$ \\
     & 2 & $0.4412223_{-0.0000022}^{+0.0000023}$ & $6.8_{-2.9}^{+1.5}$ & $-4.25_{-2.46}^{+0.61}$ & $99.59\pm 0.14$ & $1.96$ \\
     & 3 & $0.441817_{-0.000015}^{+0.000016}$ & $53.9_{-7.3}^{+8.9}$ & $-1.00_{-0.13}^{+0.13}$ & $58.95\pm 0.17$ & $1.13$ \\
     & 4 & $0.44100_{-0.00027}^{+0.00020}$ & $351_{-83}^{+131}$ & $-0.228_{-0.062}^{+0.061}$ & $6.68\pm 0.19$ & $0.97$ \\
     \hline
     2014 August 03 & 1 & $0.4412917_{-0.0000033}^{+0.0000015}$ & $4.79_{-2.43}^{+0.82}$ & $-20.5_{-20.0}^{+2.7}$ & $311.93 \pm 0.08$ & $1.96$ \\
     & 2 & $0.4412163_{-0.0000027}^{+0.0000027}$ & $8.1_{-1.1}^{+1.3}$ & $-4.64_{-0.63}^{+0.59}$ & $39.56 \pm 0.13$ & $1.32$ \\
     \hline
     2014 September 01 & 1 & $0.4412914_{-0.0000012}^{+0.0000008}$ & $4.45_{-0.53}^{+0.30}$ & $-22.2_{-2.6}^{+1.2}$ & $1188.89 \pm 0.08$ & $3.75$ \\
     & 2 & $0.4412228_{-0.0000019}^{+0.0000023}$ & $6.7_{-2.7}^{+1.3}$ & $-4.55_{-2.42}^{+0.59}$ & $139.57 \pm 0.14$ & $1.97$ \\
     & 3 & $0.441820_{-0.000017}^{+0.000017}$ & $55.7_{-7.8}^{+8.7}$ & $-0.82_{-0.11}^{+0.11}$ & $39.22 \pm 0.17$ & $1.35$ \\
     & 4 & $0.44050_{-0.00017}^{+0.00014}$ & $328_{-95}^{+119}$ & $-0.252_{-0.057}^{+0.049}$ & $17.96 \pm 0.19$ & $1.06$ \\
     \hline
     Average & 1 & $0.44129230_{-0.00000041}^{+0.00000039}$ & $4.96_{-0.16}^{+0.15}$ & $-20.38_{-0.56}^{+0.51}$ & $2417.76 \pm 0.08$ & $6.88$ \\
     & 2 & $0.4412209_{-0.0000011}^{+0.0000011}$ & $7.65_{-0.66}^{+0.64}$ & $-4.13_{-0.30}^{+0.27}$ & $309.37 \pm 0.14$ & $3.18$ \\
     & 3 & $0.441819_{-0.000010}^{+0.000010}$ & $54.2_{-5.0}^{+5.4}$ & $-0.900_{-0.075}^{+0.074}$ & $116.76 \pm 0.18$ & $1.68$ \\
     & 4 & $0.44061_{-0.00014}^{+0.00014}$ & $338_{-64}^{+73}$ & $-0.197_{-0.031}^{+0.030}$ & $24.23 \pm 0.20$ & $1.30$ \\
     \hline
   \end{tabular}
\end{threeparttable}
\end{table*}

\begin{figure*}
\centering
\includegraphics[width = 0.45\textwidth]{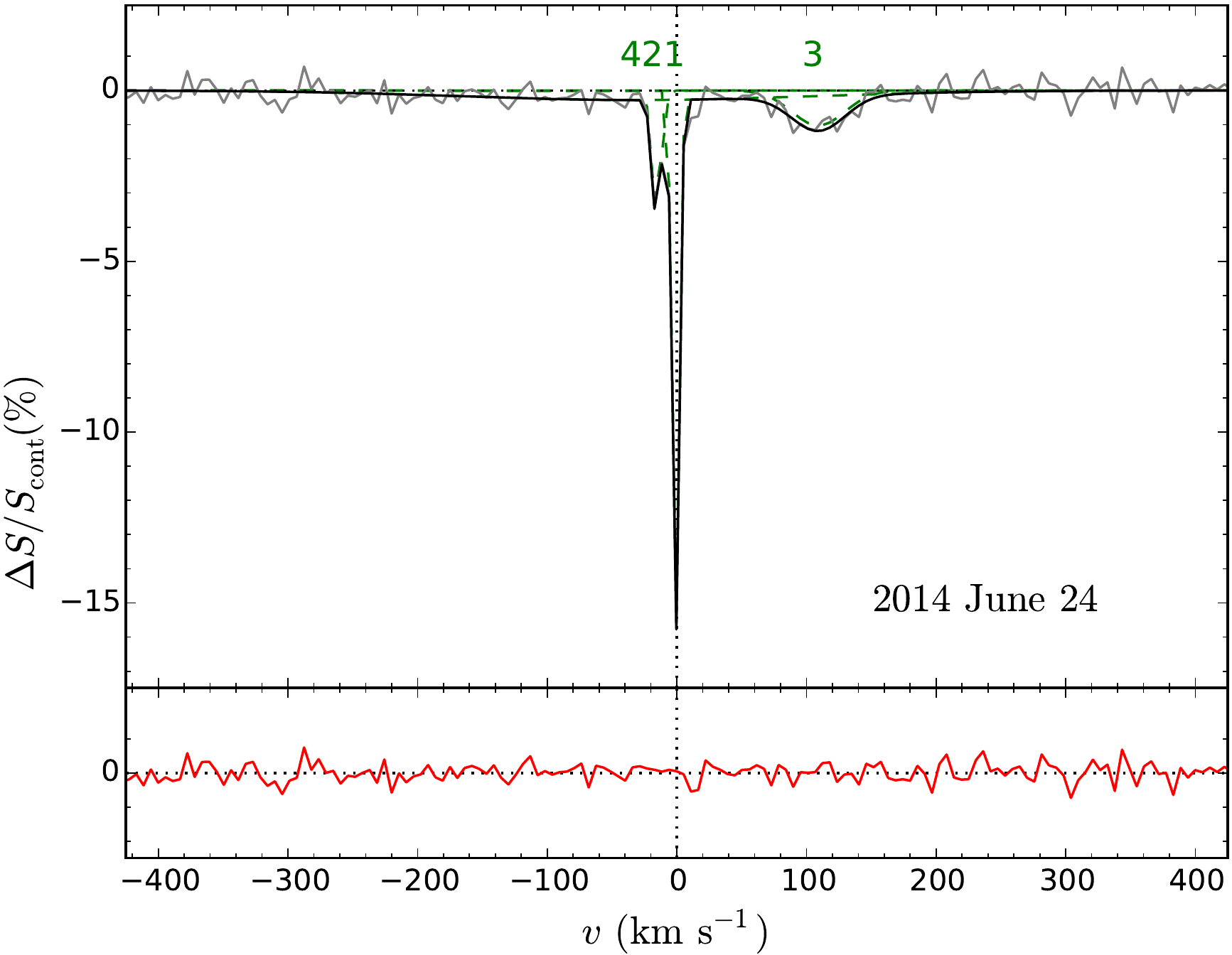}
\quad
\includegraphics[width = 0.45\textwidth]{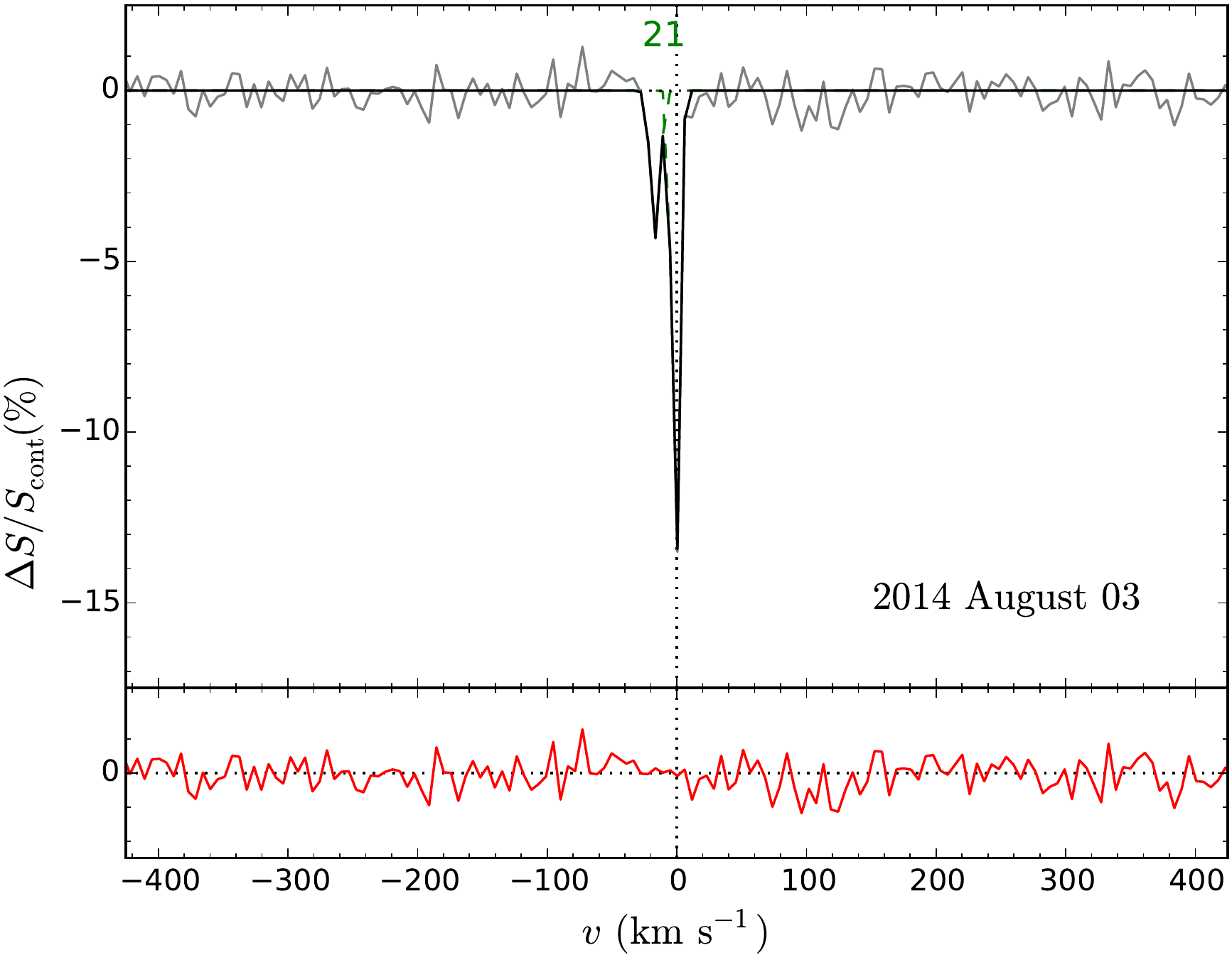} \\
\vspace{6pt}
\includegraphics[width = 0.45\textwidth]{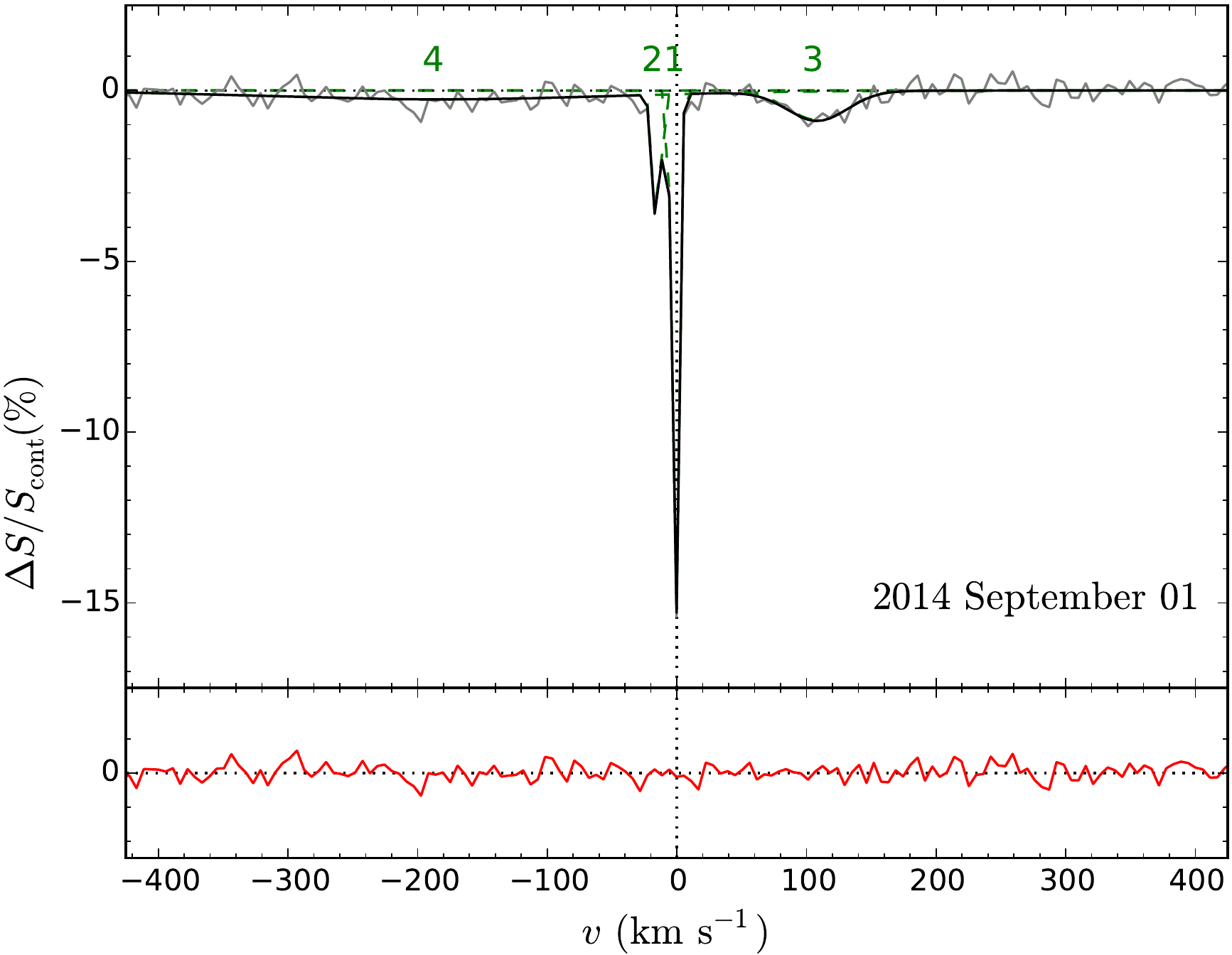} 
\quad
\includegraphics[width = 0.45\textwidth]{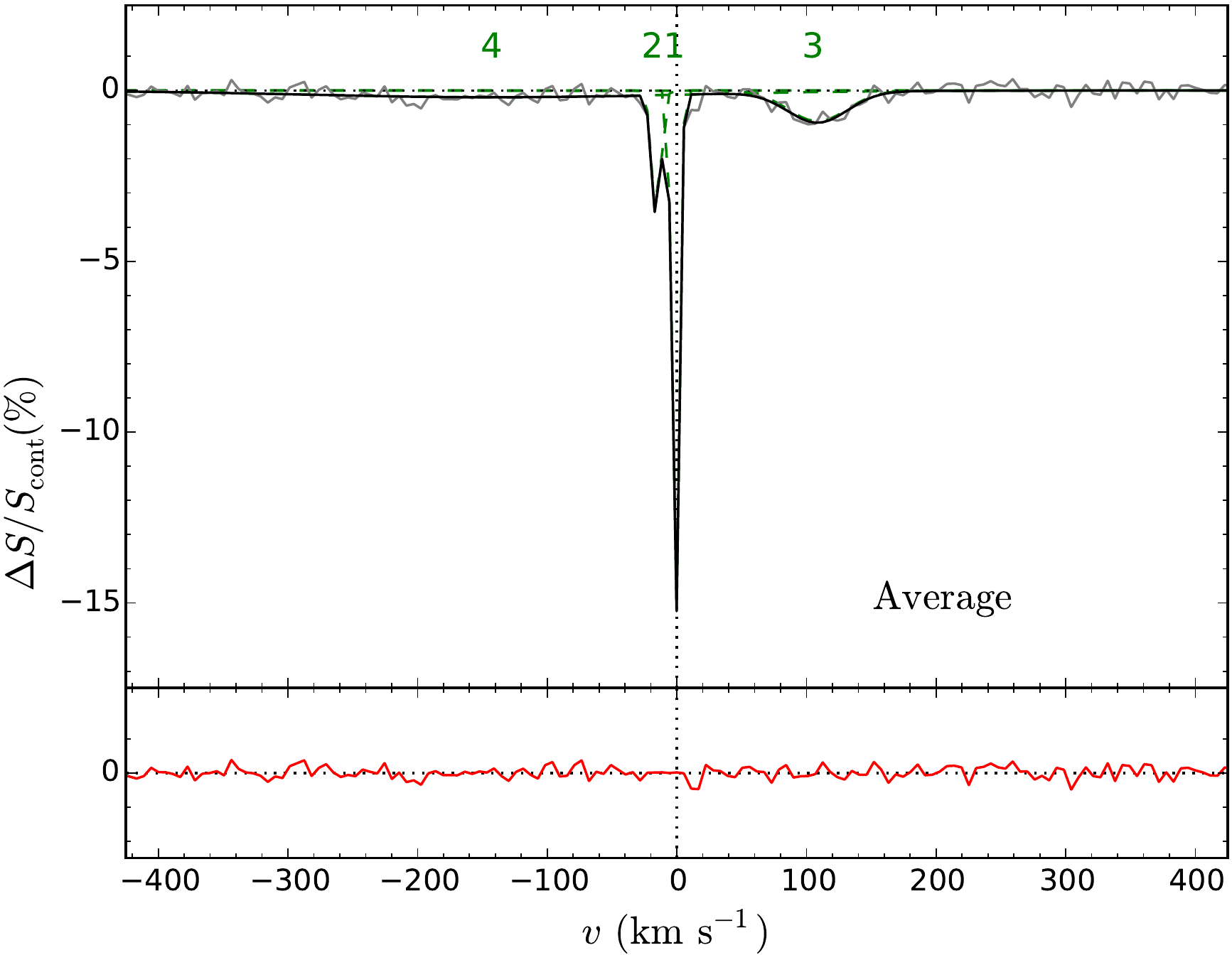} \\
\caption{The best-fitting models of the BETA spectra for each of the
  three observing epochs and the average spectrum. The radial velocity
  axis is given relative to the rest frame defined by the peak optical
  depth (vertical dotted line). The grey line represents the spectral
  data and the solid black line represents the best fitting line
  model. The dashed green lines denote the individual Gaussian
  components, which are identified by numbers that are ordered by
  descending peak line strength. The solid red line denotes the best
  fitting residual.}
\label{figure:PKS1740-517_beta_fit}
\end{figure*}

\subsection{Combining spectra from multiple PAF beams}

Some sources in the field may be common to multiple PAF beams (see
Fig.\ref{figure:PKS1740-517_beta_mosaic}) and so we construct a single
spectrum by averaging over spectra from the individual beams. Since
our arrangement of the PAF beams here is such that they do overlap,
the noise in each spectrum is not independent and so the optimal S/N
is given by weighting each spectrum by the inverse of the noise
covariance matrix. In the case of the diamond footprint used here, the
nominal noise correlation measured between adjacent beams (separated
by $1\fdg2$ between their centres) ranges from approximately 30 to
10\,per\,cent across the 711.5 -- 1015.5\,MHz band. The optimal
averaged spectrum is then given by
\begin{equation}
  \bar{S} = \sigma_{\bar{S}}^{2}\,[\bmath{w}^{\rm t}\,\mathbfss{C}_{\rm beam}^{-1}\,\bmath{S}]
\end{equation}
and
\begin{equation}
\sigma_{\bar{S}}^{2} = [\bmath{w}^{\rm t}\,\mathbfss{C}_{\rm beam}^{-1}\,\bmath{w}]^{-1},
\end{equation}
where $\mathbfss{C}_{\rm beam}$ is the noise covariance matrix between
beams, $\bmath{S}$ is the set of fractional absorption data in each
beam and $\bmath{w}$ is an accompanying set of weights, which in this
case are unity. The elements of $\mathbfss{C}_{\rm beam}$ are
estimated empirically by measuring the noise covariance between pairs
of beams on a per-pixel basis.

In Fig.\,\ref{figure:PKS1740-517_beta_spectrum} we show the resulting
full 711.5--1015.5\,MHz spectrum towards PKS\,B1740$-$517, averaged
over all three epochs. It is clear that the quality of the spectrum in
this band is extremely high and we did not need to excise any large
region due to RFI. As described above, only minor birdies were found,
and these were due to hardware issues that either have been fixed or
will not be present for the full ASKAP system. These observations make
us optimistic that ASKAP will be a successful instrument in this
frequency range.

\section{Results}

\subsection{HI absorption towards PKS\,B1740$-$517}

We search for the signature of \mbox{H\,\sc{i}} absorption in the BETA
data using a spectral line detection and modelling technique based on
Bayesian model comparison (see \citealt{Allison:2014} and references
therein). This technique is a robust method of model selection,
enabling us to both assign a significance to the detection and compare
between increasingly complex model representations of the data. The
spectral data are modelled by convolving a physical model of the line,
consisting of multiple Gaussian components, with the known gain
response of the fine filterbank channels. Using this method we detect
a single absorption line in the 21-cm spectrum of PKS\,B1740$-$517,
located at an observed frequency of 985.5\,MHz, equal to a
\mbox{H\,{\sc i}} redshift of $z = 0.4413$. By sequentially increasing
the number of Gaussian components, and comparing the Bayes factors, we
arrive at a best-fitting spectral model for each of our observing
epochs, and for the average spectrum. In
Table\,\ref{table:model_parameters} we summarize the model parameters
and in Fig.\,\ref{figure:PKS1740-517_beta_fit} we show the best
fitting Gaussian components and the total model.

We find that the combined data are best fitted by four components,
corresponding to two narrow components (widths $\sim$5 and
8\,km\,s$^{-1}$) at the position of peak absorption and two broad
components (widths $\sim$50 and 350\,km\,s$^{-1}$), which are,
respectively, redshifted and blueshifted with reference to the peak
absorption. The latter blueshifted broad component, with a peak at
approximately 0.2\,per\,cent of the continuum, is arguably the most
tentative.  However, we believe the veracity of this detection for the
following reasons: (a) the inclusion of this component is strongly
warranted by the data above the formal noise, with an increase in log
evidence of $\Delta{\ln(\mathcal{Z})} = +24.23 \pm 0.20$ and a
decrease in reduced chi-squared for the best-fitting model parameters
of $\Delta{\chi^{2}/{\rm d.o.f.}} = -0.38$; (b) other such features
are not present elsewhere, either in this spectrum or other sources in
the field (see Section\,\ref{section:other_sources}); (c) the feature
is stable to changes in our continuum subtraction procedure. While
formally detected above the noise, the peak strength of this feature
is weak and further observations are needed for confirmation. We
discuss the possible physical interpretations of these results in
Section\,\ref{section:discussion}.

\subsection{Spectral variability}

Our observations with BETA were carried out in three separate epochs
with adjacent intervals of 40 and 29\,d. Over these time intervals we
can test for variability of the individual spectral
components. Limited observational evidence in the literature indicates
that $\sim10$\,per\,cent stochastic fluctuations may occur in the
relative fluxes of some \mbox{H\,{\sc i}} absorption components on
time-scales of days to several weeks; prominent examples are the two
intervening absorption systems towards the flat-spectrum quasars
AO\,0235+164 (\citealt{Wolfe:1982}) and PKS\,B1127$-$145
(\citealt{Kanekar:2001}), which exhibit flux variation but no
significant shift in position or width. Several possible models exist
to explain such behaviour, which include motion of radio source
components with respect to the foreground absorber
(\citealt{Briggs:1983}), interstellar scintillation in the Milky Way
Galaxy (\citealt{Macquart:2005}), and microlensing of the background
source (\citealt{Lewis:2003}). Due in part to limited channel sampling
of the strongest absorption components, our data are sensitive (at
3\,$\sigma$) to fractional variations in the optical depth of only
30\,per\,cent over a 2\,month interval. For a background source
traversing the absorber at a velocity close to the speed of light, our
data are sensitive to fluctuations in the 21-cm opacity of more than
30\,per\,cent on transverse scales less than 0.05\,pc. If we consider
instead the case of interstellar scintillation, for an absorber at $z
= 0.44$, the required physical scales are much less than a few
parsecs. Since we find no significant evidence for variability in the
component parameters given in Table\,\ref{table:model_parameters} this
suggests that the bulk of the cold \mbox{H\,{\sc i}} gas in these
components is distributed uniformly over parsec scales, consistent
with that found in Local Group galaxies (c.f. $\sim$\,100\,pc;
\citealt{Braun:2012}) and radio galaxies (\citealt{Curran:2013b}).
Given our time sampling of approximately a month, it is also possible
that significant scintillation may occur on day and intraday
time-scales to which our data are not yet sensitive.

\subsection{Intervening absorption towards other
  sources}\label{section:other_sources}

\begin{figure}
\centering
\includegraphics[width = 0.45\textwidth]{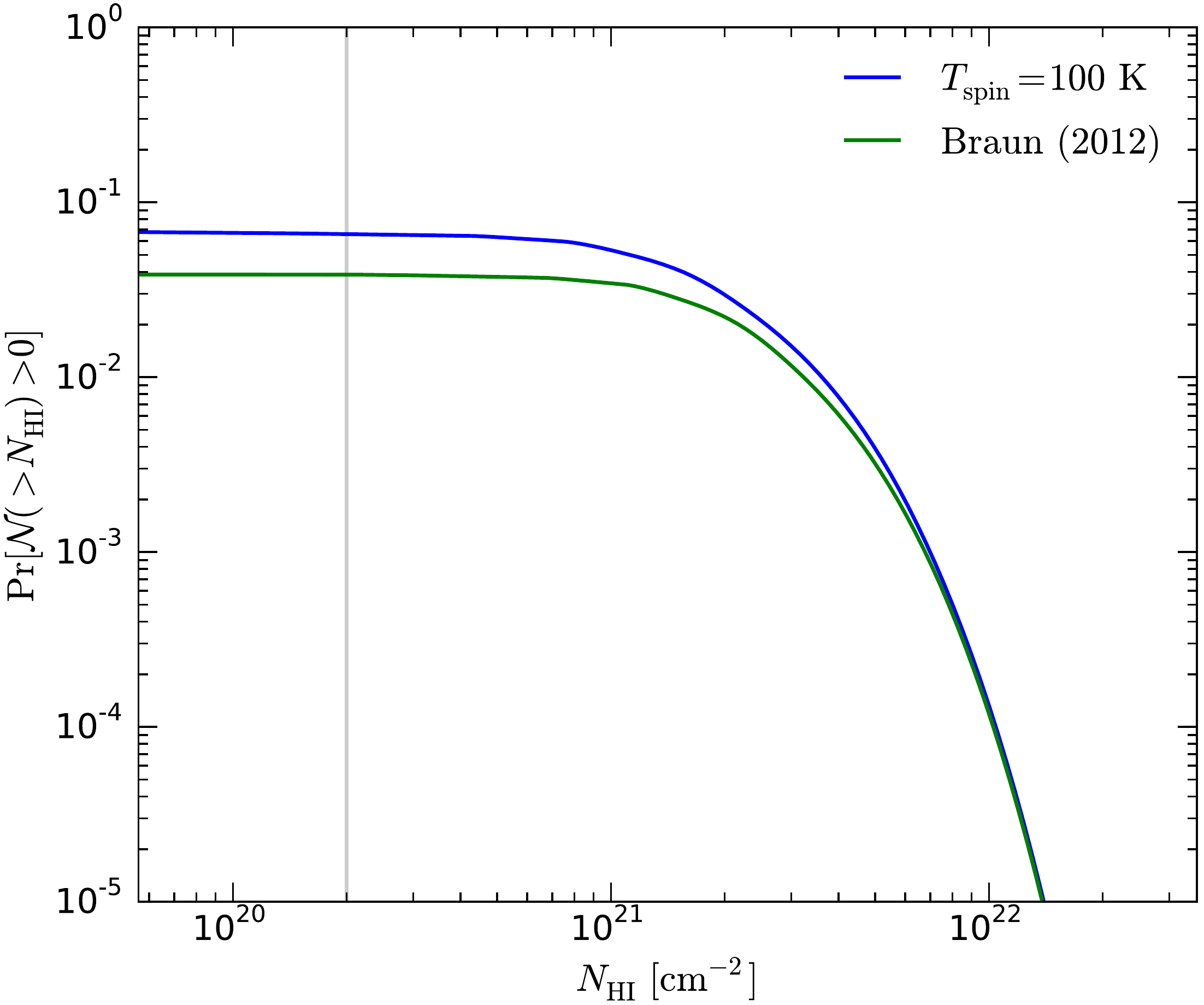}
\caption{The probability of detecting at least one intervening
  absorber with column density greater than $N_{\rm HI}$ in the
  PKS\,B1740$-$517 field observed with BETA. The blue line represents
  a simple Gaussian model for the \mbox{H\,{\sc i}} gas, assuming that
  $c_{\rm f} = 1$, a spin temperature of $T_\rmn{spin} = 100$\,K and
  $\Delta{v_{50}} = 30$\,km\,s$^{-1}$. The green line represents the
  two-phase temperature-shielding model by \citet{Braun:2012}, with
  $N_{0} = 1.25\times10^{20}$\,cm$^{-2}$ and $N_{\infty} =
  7.5\times10^{21}$\,cm$^{-2}$. The vertical grey line denotes a
  Damped Ly$\alpha$ absorber ($N_{\rm HI} > 2 \times
  10^{20}$\,cm$^{-2}$), showing that with BETA we are only sensitive
  to systems with significantly higher column densities.}
\label{figure:nhi_distribution}
\end{figure}

The field of view available with nine PAF beams enables a wider search
for intervening \mbox{H\,{\sc i}} absorption in the spectra of
multiple sources. Based on a nominal noise per channel of
20\,mJy\,beam$^{-1}$ for the data averaged over all three
observations, we searched the spectra of 72 sources brighter than
100\,mJy within 1\degr\ of each beam centre, selected from the SUMSS
(\citealt{Mauch:2003}) and MGPS-2 (\citealt{Murphy:2007}) catalogues,
giving a 5\,$\sigma$ detection limit for absorption against the
weakest sources at the beam centre. Using the spectral-line finding
technique of \cite{Allison:2014} we found no further \mbox{H\,{\sc i}}
absorption in the spectra of radio sources in this field i.e. all
spectra are consistent with the noise. 

We can estimate the expected number of intervening absorbers $n$ with
a column density greater than $N_{\rm HI}$ by
\begin{eqnarray}\label{equation:expected_number}
  n(>N_{\rm HI}) & = & \int_{N_{\rm HI}}^{\infty}\int{f(N_{\rm HI}^{\prime},X)\,\mathrm{d}X\,\mathrm{d}N_{\rm HI}^{\prime}} \nonumber \\ 
  & \approx & \sum_{N_{\rm HI}}^{\infty}\sum{f(N_{\rm HI}^{\prime},X)\,\delta{X}\,\delta{N}_{\rm HI}^{\prime}},
\end{eqnarray}
where $f$ is the frequency distribution of systems with column density
$N_{\rm HI}$ and $X$ is the comoving path length given (in a flat
$\Lambda$CDM universe) by
\begin{equation}
  X(z) = {2 \over 3\Omega_{\rm M}}\sqrt{\Omega_{\rm M}(1+z)^{3} + \Omega_\Lambda}.
\end{equation} 
To estimate $f$ as a function of redshift we perform a simple linear
interpolation between distributions measured for the local Universe
(\citealt{Zwaan:2005}) and at $z = 3$ (\citealt{Noterdaeme:2009}).

For each sight line towards our 72 sources we calculate the comoving
path interval by
\begin{equation}
    \delta{X}= 
\begin{cases}
  X(z+0.5\,\delta{z})-X(z-0.5\,\delta{z}), & \text{if}\ N_{\rm HI}\prime > N_{5\sigma}, \\
  0, & \text{otherwise},
\end{cases}
\end{equation}
where $N_{5\sigma}$ is the column density sensitivity for detection of
an absorption line as a function of spectral channel, and $\delta{z}$
is the channel separation in redshift. Given that the BETA spectra
span a wide redshift interval, and that absorption can only be
detected between the source and observer, we must also take into
consideration the distribution of source redshifts. We therefore use
the following redshift distribution model of \cite{deZotti:2010}
\begin{equation}
  \mathcal{N}_{\rm s}(z) = 1.29 + 32.37z - 32.89z^{2} + 11.13z^{3} - 1.125z^{4},
\end{equation}
which is determined from the population of radio sources brighter than
10\,mJy at 1.4\,GHz. For any given sight line, the probability of the
source being located at a redshift $z_{\rm s}$ beyond $z$ is equal to
\begin{equation}
  \mathrm{Pr}(z_{\rm s}>z) = {\int_{z}^{\infty} \mathcal{N}_{\rm s}(z^{\prime})\mathrm{d}z^{\prime}\over\int_{0}^{\infty}\mathcal{N}_{\rm s} (z^{\prime})\mathrm{d}z^{\prime}},
 \end{equation}
 and so the expected value for the absorption comoving path length
 probed by each spectral channel is given by
 \begin{equation}
   \langle{\delta{X}(z)}\rangle = \mathrm{Pr}(z_{\rm s}>z)\delta{X}(z).
 \end{equation}

 We consider two models that convert the observed spectrum (in units
 of fractional absorption) to the column density sensitivity. First we
 use a simple conversion assuming line-of-sight gas of spin
 temperature $T_{\rm spin} = 100$\,K, covering factor $c_{\rm f} = 1$
 and velocity FWHM $\Delta{v_{50}} = 30$\,km\,s$^{-1}$. The column
 density sensitivity is then given by
\begin{equation}\label{equation:model_one}
  N_{5\sigma} = 5.82\times10^{21}\left[{T_{\rm spin}\over 100\,\mathrm{K}}\right]\left[{\Delta{v_{\rm 50}}\over 30\,\mathrm{km}\,\mathrm{s}^{-1}}\right]\tau_{\,5\sigma}\,\mathrm{cm}^{-2},
\end{equation}
where $\tau_{\,5\sigma}$ is the optical depth sensitivity estimated
from the spectral noise per channel and assuming a covering factor of
unity. In the case of our second model we use a simple two-phase
sandwich geometry for the gas where the cold neutral medium (CNM;
$T_{\rm c} \sim 100$\,K) is sandwiched by a layer of warm neutral
medium (WNM; $T_{\rm w} \gtrsim 5000$\,K), which acts as a shield from
local high energy UV and X-ray radiation (for further details see
\citealt{Kanekar:2011} and \citealt{Braun:2012}). In this case the
column density sensitivity is given by
\begin{equation}\label{equation:model_two}
  N_{5\sigma} = N_{0} + \left[N_{\infty} - \left({N_{0} \over 2}\right)\right]\tau_{\,5\sigma},
\end{equation}
where $N_{0}$ is the threshold column density of WNM at which the CNM
can form, and $N_{\infty}$ is the saturation column density as the
optical depth in CNM tends to infinity. We use values for these
parameters of $N_{0} = 1.25 \times 10^{20}$\,cm$^{-2}$ and $N_{\infty}
= 7.5 \times 10^{21}$\,cm$^{-2}$, obtained by \cite{Braun:2012} from
model fitting to detailed 21-cm observations of the Milky Way Galaxy,
Messier\,31 and the Large Magellanic Cloud.

The probability of detecting $\mathcal{N}$ absorbers with column
densities greater than $N_{\rm HI}$ is then given by
\begin{equation}
  \mathrm{Pr}[\mathcal{N}(>N_{\rm HI})] = {n^{\mathcal{N}} \over \mathcal{N}!} \exp(-n),
\end{equation}
where $n$ is the expected number of absorbers given by
Equation\,\ref{equation:expected_number}. Based on the above models of
the \mbox{H\,{\sc i}} gas we expect to detect less than $\sim 0.05$
intervening absorption systems with column densities greater than
about $10^{21}$\,cm$^{-1}$, with a corresponding probability of less
than $\sim 0.05$ for detecting at least one such system (see
Fig.\,\ref{figure:nhi_distribution}). Therefore the lack of any
intervening detection in our data is consistent with the known
distributions of \mbox{H\,{\sc i}} in the local and distant Universe.

\section{Multi-wavelength follow-up}

\begin{figure*}
\centering
\includegraphics[width = 0.95\textwidth]{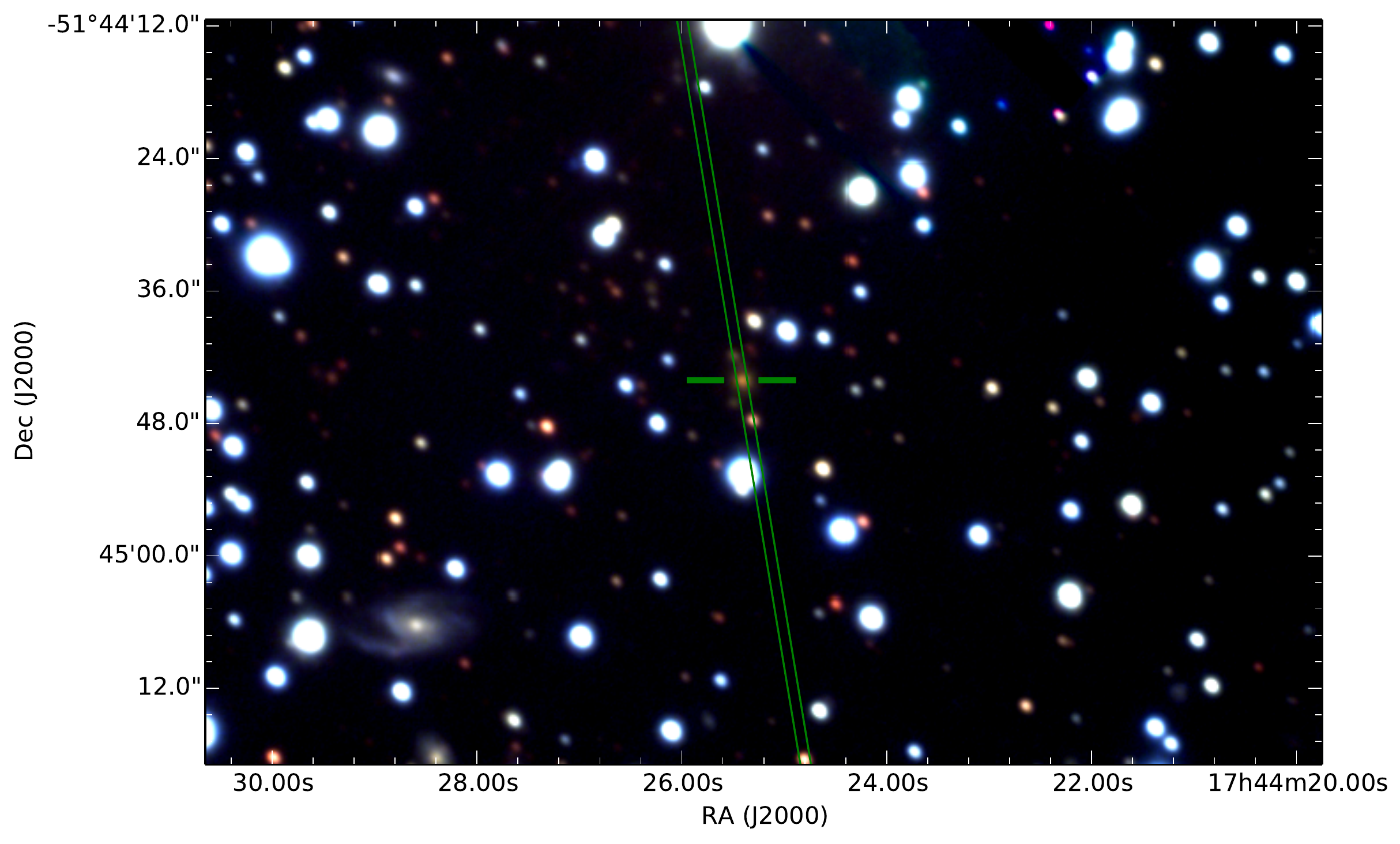}
\caption{A three-colour optical image constructed from $g'r'i'$-band
  observations with the 8-m Gemini-South telescope. PKS\,B1740$-$517
  is identified with the red galaxy located between the two solid
  horizontal bars. The orientation and position of the 1.5-arcsec slit
  are indicated by the parallel lines.}
\label{figure:PKS1740-517_gemini_image}
\end{figure*}

\begin{figure*}
\centering
\includegraphics[width = 0.95\textwidth]{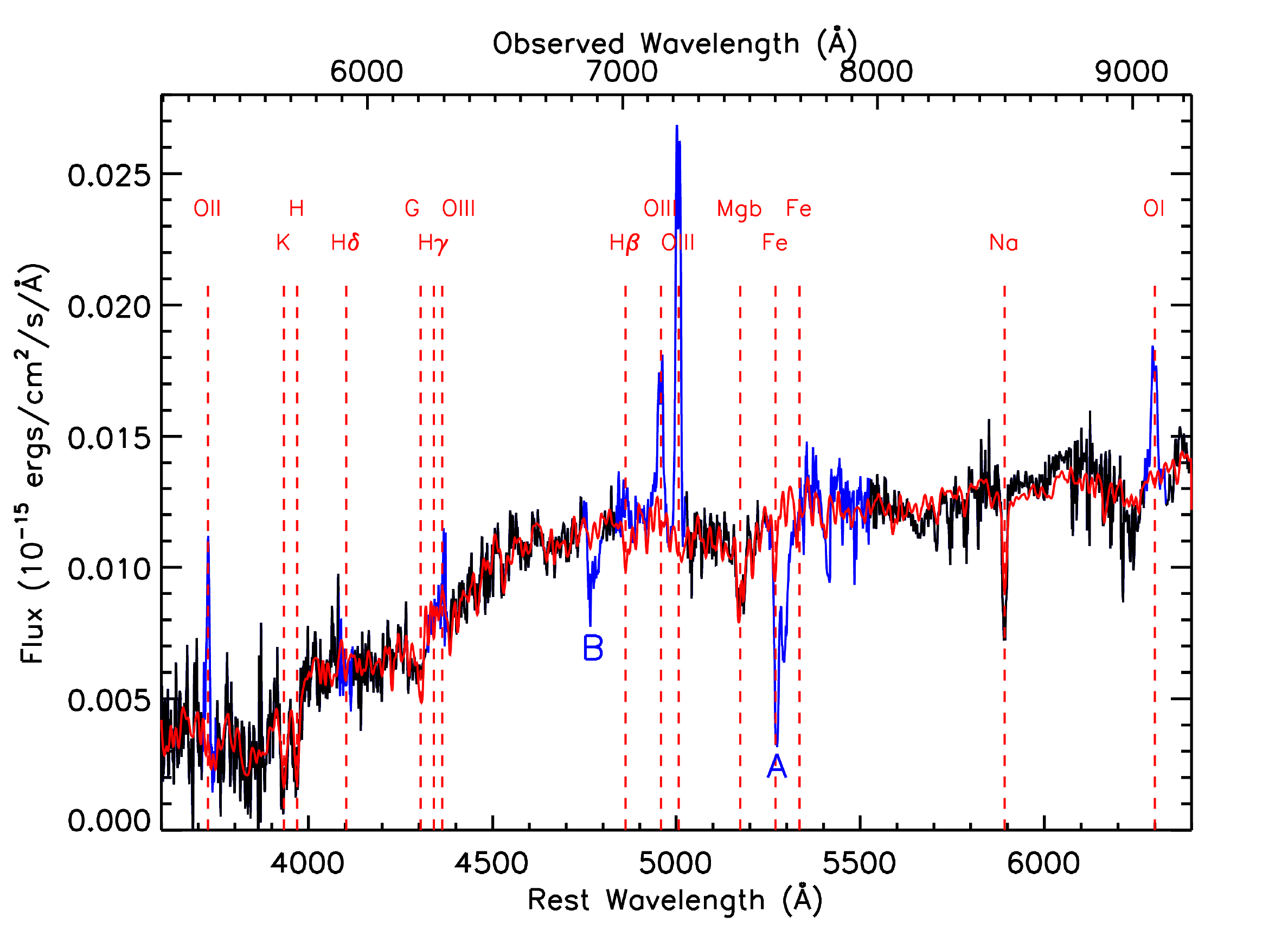}
\caption{The optical spectrum towards PKS\,B1740$-$517 from long-slit
  observations using the GMOS instrument on the Gemini-South
  Telescope. The black line denotes the full spectrum and the vertical
  dotted lines denote the rest wavelengths of known species. We have
  performed a model fit to the stellar population (red line) using
  \texttt{pPXF}, where the emission lines and Telluric regions (blue
  lines) have been masked, obtaining a systemic redshift of $z_{\rm
    sys} = 0.44230 \pm 0.00022$.}
\label{figure:PKS1740-517_gemini_spectrum}
\end{figure*}

\subsection{Optical spectroscopy with Gemini-South}

\subsubsection{Observations and data analysis}

\begin{table} 
  \begin{threeparttable}\setlength\tabcolsep{3pt}\renewcommand*\arraystretch{1.3}
    \caption{Summary of emission line properties from our Gemini-South
      observations.}\label{table:emission_lines}
  \begin{tabular}{lccc} 
    \hline
    \multicolumn{1}{l}{Line} & \multicolumn{1}{c}{$\log$(flux)} & \multicolumn{1}{l}{Equivalent width} & \multicolumn{1}{l}{$\log$(luminosity)} \\
    \multicolumn{1}{c}{} & \multicolumn{1}{c}{(erg\,s$^{-1}$\,cm$^{-2}$)} & \multicolumn{1}{c}{(\AA)} & \multicolumn{1}{c}{(erg\,s$^{-1}$)} \\   
    \hline
    \mbox{[O\,{\sc ii}]} $\lambda3727$ & $-16.11_{-0.04}^{+0.04}$ & $23.60\pm2.26$ & $40.74_{-0.04}^{+0.04}$ \\
    \mbox{H$\beta$} & $-16.34_{-0.12}^{+0.09}$ & $4.22\pm1.00$ & $40.52_{-0.12}^{+0.09}$ \\
    \mbox{[O\,{\sc iii}]} $\lambda 4959$ & $-16.00_{-0.05}^{+0.04}$ & $8.52\pm0.91$ & $40.86_{-0.05}^{+0.04}$ \\
    \mbox{[O\,{\sc iii}]} $\lambda 5007$ & $-15.58_{-0.02}^{+0.02}$ & $23.73\pm0.95$ & $41.27_{-0.02}^{+0.02}$ \\
    \mbox{[O\,{\sc i}]} $\lambda 6300$ & $-16.07_{-0.07}^{+0.06}$ & $6.68\pm0.96$ & $40.78_{-0.07}^{+0.06}$ \\
    \hline
   \end{tabular}
\end{threeparttable}
\end{table}

As discussed in Section\,\ref{section:pks1740-517}, prior to our
observations with the BETA telescope the redshift of the optical
counterpart to PKS\,B1740$-$517 was not well known, with no
spectroscopically determined value available from the
literature. Furthermore, the \mbox{H\,{\sc i}} redshift did not match
any of the published photometric estimates. The interpretation of the
observed absorption required an accurate knowledge of the host galaxy
redshift, so that the \mbox{H\,{\sc i}} gas could be associated either
with the host galaxy of the radio source or with an intervening
system.

We therefore obtained time from Gemini-South under the Director's
Discretionary Time program (proposal code GS-2014B-DD-2) to acquire
long-slit spectroscopy and $g'r'i'$-band imaging with the Gemini
Multi-Object Spectrograph (GMOS; \citealt{Hook:2003}). The long-slit
spectroscopy was taken with the R400 grating, utilizing two grating
settings with central wavelengths 7000 and 7050\,\AA. This was chosen
to put the \mbox{H$\beta$} and \mbox{[O\,{\sc iii}]} lines in the
centre of the spectrum for a host redshift equal to that of the
\mbox{H\,{\sc i}} absorption. At each grating setting we obtained
$3\times900$\,s exposures, giving a total exposure time of 1.5\,h. A
1.5-arcsec slit was used, and the CCD was binned $2\times2$ on-chip.

The spectra were reduced in \texttt{IRAF}, using standard techniques
within the Gemini package. The spectra were bias-subtracted and
flat-fielded using calibration frames from the Gemini Facility
Calibration Unit (GCAL). Wavelength calibration made use of Cu-Ar
comparison spectra, and the spectrum was flux calibrated using
observations of the standard star Feige 110. We estimate an rms error
of 0.9\,\AA\ in the wavelength calibration, based on the spread in the
Cu-Ar lines. We also acquired images with GMOS in the $g'r'i'$
bands. Each image consisted of $4 \times 75$\,s exposures, which were
binned $2 \times 2$ on-chip, giving a pixel scale of 0.16
arcsec\,pixel$^{-1}$. A three-colour image ($g'$ as blue, $r'$ as
green and $i'$ as red) is shown in
Fig.\,\ref{figure:PKS1740-517_gemini_image}. Adjusting the original
astrometry of the Gemini image using the $B_{\rm j}$-band image from
the SuperCosmos Sky Survey (\citealt{Hambly:2001}) we find that the
centroid position of the large central galaxy (indicated by the solid
horizontal bars in Fig.\,\ref{figure:PKS1740-517_gemini_image}) agrees
with the ICRF2 position of PKS\,B1740$-$517 to within 0.6\,arcsec.

The full GMOS spectrum is shown in
Fig.\,\ref{figure:PKS1740-517_gemini_spectrum} with a measured FWHM
spectral resolution of 11\,\AA\ from fitting the Cu-Ar arc
spectra. Clearly apparent are the \mbox{[O\,{\sc iii}]}
$\lambda\lambda 4959,5007$ doublet, the \mbox{[O\,{\sc ii}]}
$\lambda3727$ and \mbox{[O\,{\sc i}]} $\lambda6300$ emission lines,
all consistent with a redshift that matches the \mbox{H\,{\sc i}}
absorption. There is also evidence of \mbox{[O\,{\sc iii}]}
$\lambda4363$ emission. However, it is contaminated by subtraction of
the 6300\,\AA~sky line and so we cannot infer accurate kinematics for
this line. We note that there is almost no sign of Balmer-line
emission, with weak \mbox{H$\beta$} at $4861$\,\AA~and
\mbox{H$\alpha$} redshifted out of the observed band. In order to
obtain an estimate of the systemic redshift of the host galaxy, we use
the Penalized Pixel-Fitting method (\texttt{pPXF};
\citealt{Cappellari:2004}) to fit stellar population synthesis models
(\citealt{Vazdekis:2010}) to the underlying stellar continuum. Using
this method we obtain a systemic redshift of $z = 0.44230 \pm
0.00022$, where the uncertainty is derived by summing in quadrature
the formal error due to the noise ($\sigma_{z,\mathrm{noise}} =
0.00013$) and the wavelength calibration ($\sigma_{z,\mathrm{cal}} =
0.00018$). We therefore conclude that the \mbox{H\,{\sc i}} absorption
system seen towards PKS\,B1740$-$517 is intrinsic to the galaxy
hosting the radio AGN.

\subsubsection{Emission line ratios and kinematics}

\begin{figure}
\centering
\includegraphics[width = 0.5\textwidth]{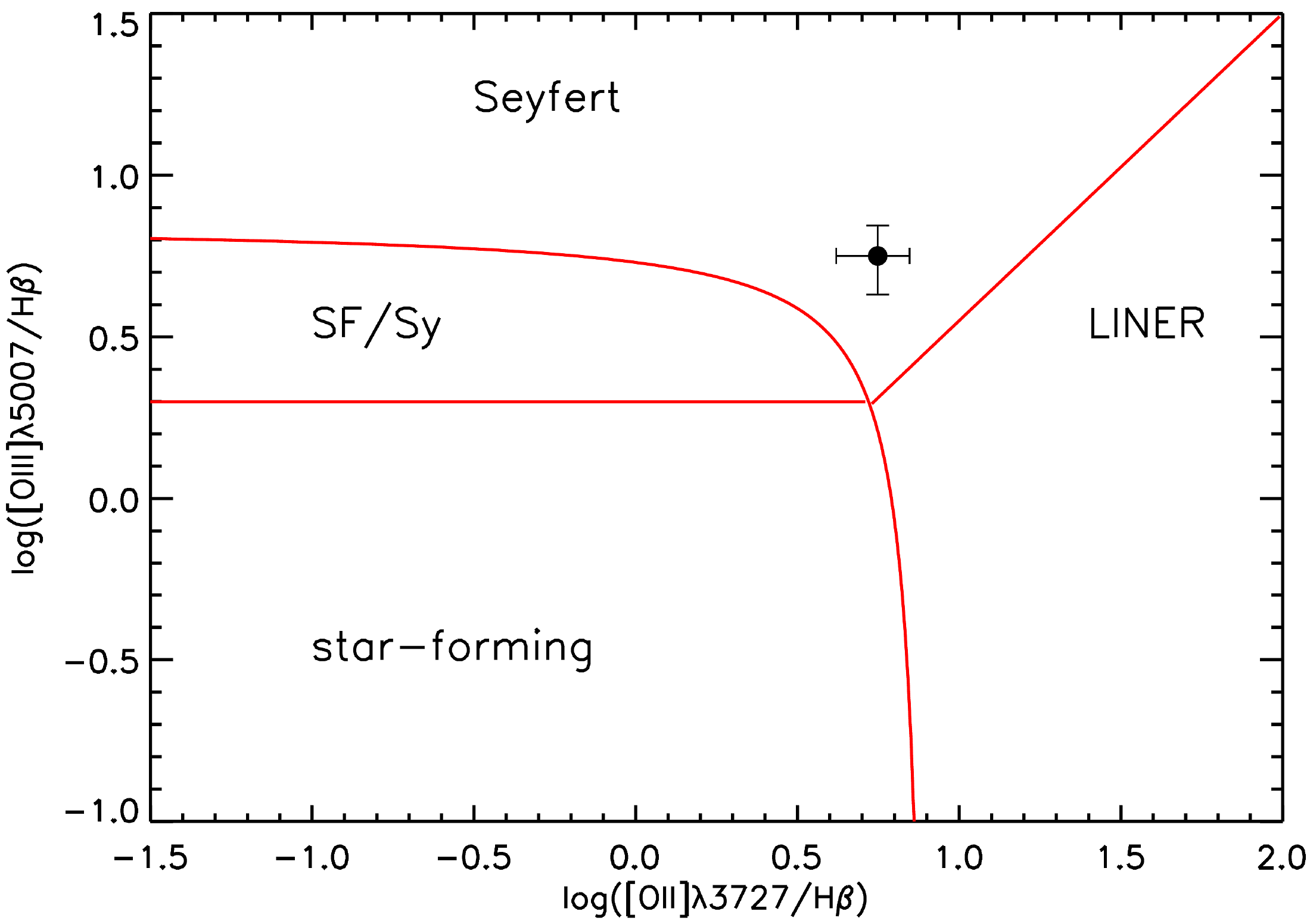}
\caption{The blue classification diagram for redshifted emission line
  galaxies (\citealt{Lamareille:2010}), where the position of
  PKS\,B1740$-$517 is indicated by the error bars.}
\label{figure:PKS1740-517_blue_diagram}
\end{figure}

In Table\,\ref{table:emission_lines} we summarize the properties of
individual emission lines seen in the Gemini spectrum, where the
fluxes and equivalent widths are estimated from fitting Gaussian
models to the emission lines after subtraction of the best-fitting
stellar population models. The continuum level, used in calculating
the equivalent width, is measured adjacent to the relevant line from
the original unsubtracted spectrum. Equivalent width ratios of
\mbox{[O\,{\sc iii}]}$\lambda5007$/\mbox{H$\beta$} = $5.60 \pm 1.43$
and \mbox{[O\,{\sc ii}]}$\lambda3727/$\mbox{H$\beta$} = $5.64 \pm
1.34$ place PKS\,B1740$-$517 in the Seyfert region of the blue
classification diagram for redshifted emission line galaxies
(Fig.\,\ref{figure:PKS1740-517_blue_diagram};
\citealt{Lamareille:2010}). The \mbox{[O\,{\sc i}]} $\lambda 6300$
line is particularly strong (\mbox{[O\,{\sc i}]}$\lambda
6300$/\mbox{[O\,{\sc iii}]}$\lambda5007$ = $0.28 \pm 0.04$), which in
combination with the parity in strength of the \mbox{[O\,{\sc iii}]}
$\lambda5007$ and \mbox{[O\,{\sc ii}]} $\lambda3727$ lines could be an
indicator of Low-Ionization Nuclear Emission line Region (LINER;
\citealt{Heckman:1980}) behaviour. Strong \mbox{[O\,{\sc i}]} $\lambda
6300$ and \mbox{[O\,{\sc iii}]} $\lambda4363$ emission arise from
excitation by fast shocks (300 -- 500\,km\,s$^{-1}$), which also
indicates possible interaction between the radio jets and interstellar
gas (\citealt{Dopita:1995}).

\begin{figure}
\centering
\includegraphics[width = 0.5\textwidth]{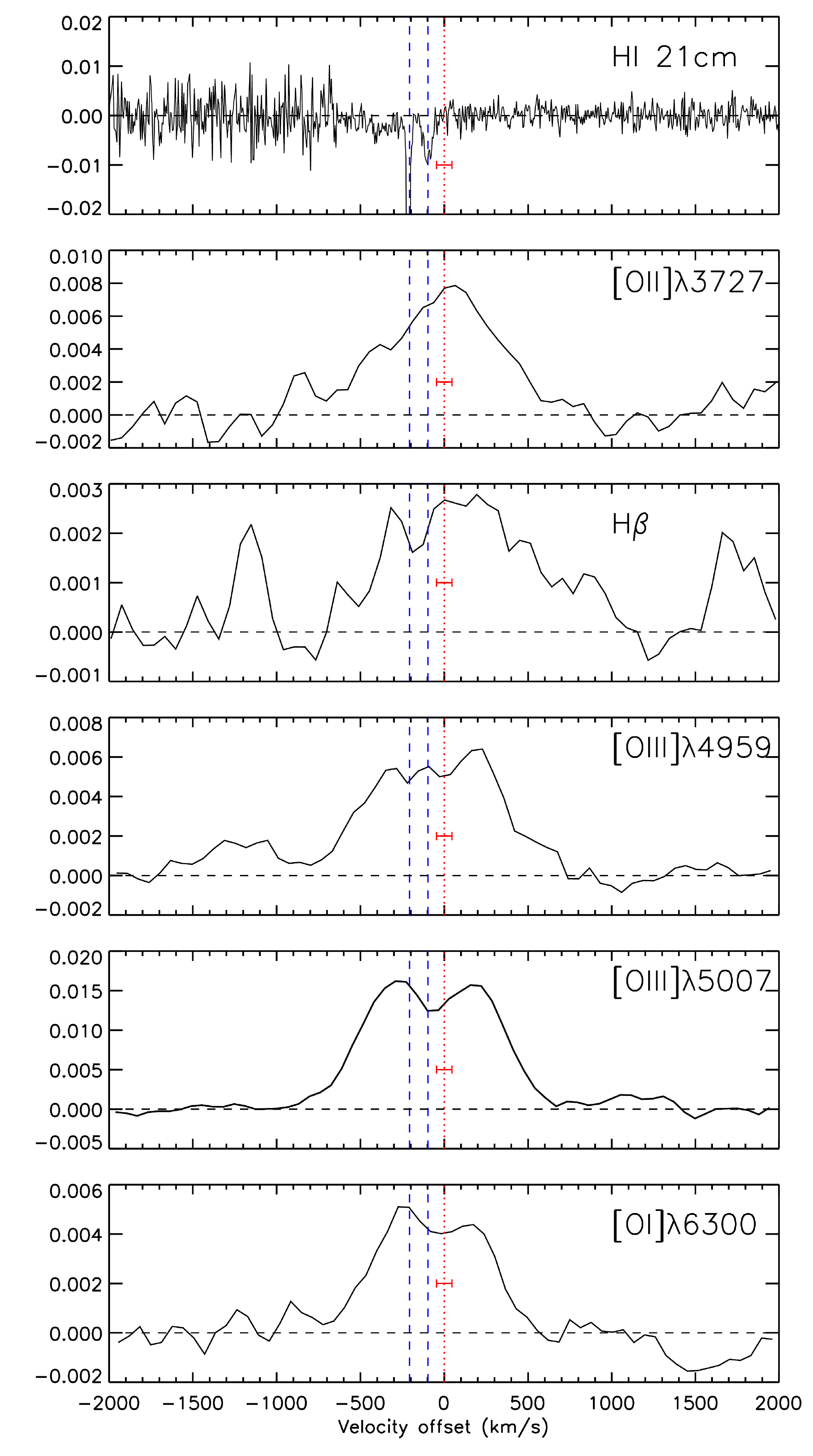}
\caption{The \mbox{H\,{\sc i}} 21-cm absorption, oxygen and H$\beta$
  emission lines in the rest frame defined by the systemic redshift of
  the galaxy ($z_{\rm sys} = 0.44230 \pm 0.00022$). In all cases the
  continuum has been subtracted from the spectrum. The vertical dashed
  lines denote the location of \mbox{H\,{\sc i}} absorption components
  1 (left) and 3 (right), which correspond to the most prominent
  features. The red horizontal error bar indicates the
  $\pm46$\,km\,s$^{-1}$ uncertainty in the systemic velocity.}
\label{figure:PKS1740-517_line_comparison}
\end{figure}

In Fig.\,\ref{figure:PKS1740-517_line_comparison} we show the velocity
structure of the optical emission lines (after subtraction of the
fitted stellar component) in the rest frame defined by the systemic
redshift. Despite the low spectral resolution (ranging from 350 to
650\,km\,s$^{-1}$ across the observed band), there is strong evidence
for double-peaked structure in the \mbox{[O\,{\sc iii}]} and
\mbox{[O\,{\sc i}]}~$\lambda6300$ emission lines. In particular, the
high S/N \mbox{[O\,{\sc iii}]}~$\lambda5007$ line clearly shows two
peaks separated by a velocity of approximately
500\,km\,s$^{-1}$. Conversely, the strong \mbox{[O\,{\sc
    ii}]}~$\lambda3727$ line appears to have only a single peak at the
systemic velocity, but is at much lower S/N than the two
\mbox{[O\,{\sc iii}]} lines and is likely confused by the unresolved
doublet.

\begin{figure}
\centering
\includegraphics[width =
0.45\textwidth]{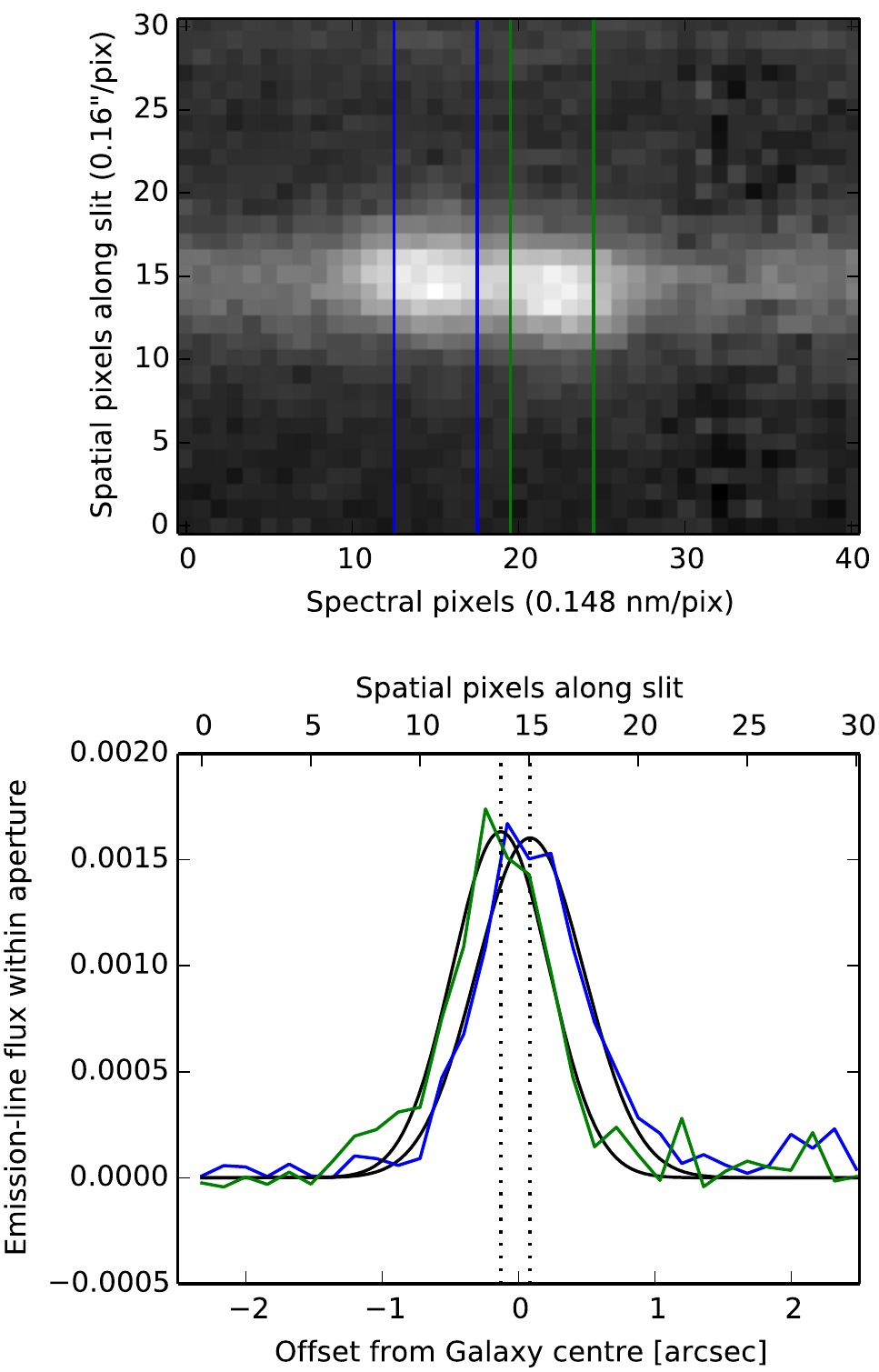}
\caption{Top: an extract of the two-dimensional spectrum showing just
  the \mbox{[O\,{\sc iii}]} $\lambda5007$ line. The axes show the
  pixels from the arbitrary start of the cut-out. The spatial axis
  pixels increase in a roughly northerly direction (consult
  Fig.\,\ref{figure:PKS1740-517_gemini_image} for the orientation of
  the slit on the sky). The two apertures used to examine the spatial
  profile of the emission line are shown, one for each peak of the
  $\lambda5007$ line. Bottom: the spatial profile averaged over each
  aperture, after subtracting the average profile across the entire
  cut-out. The colour of the profile matches the aperture shown in the
  2D spectrum. Also plotted (in black) are Gaussian fits to the
  profile used to estimate the peak location for each pixel. The
  profiles are plotted as a function of distance from the peak of the
  average profile, with the pixel values indicated for comparison
  (positive offsets are in a roughly northerly direction).}
\label{figure:PKS1740-517_OIII_position}
\end{figure}

The relative strength of the narrow oxygen and \mbox{H$\beta$} lines
strongly indicate that the emission is dominated by gas ionized by the
AGN, rather than star formation. So far only a few AGN with
double-peaked \mbox{[O\,{\sc iii}]} lines have been reported in the
literature \citep{Liu:2010,Smith:2010,Shen:2011}, and their origin has
largely been ascribed to kinematic effects within a single, rather
than binary, AGN (see \citealt{Fu:2012}). In
Fig.\,\ref{figure:PKS1740-517_OIII_position} we show the spatial
distribution of the \mbox{[O\,{\sc iii}]}~$\lambda5007$ line along the
slit, which exhibits a clear separation of 0.22\,arcsec (1.3\,kpc at
$z = 0.44$) between the peaks. Such behaviour could be due to an
outflow of AGN-driven gas (e.g \citealt{Fischer:2011}), perhaps even
directly as a result of radio jet--cloud interaction
(e.g. \citealt{Stockton:2007,Rosario:2010}), but could also be due to
kpc-scale rotation of the ionized gas. While detection of an ionized
outflow would be typical of the host galaxy of a powerful compact
radio source, the large scale seen here would be at odds with other
GPS sources, where ionized outflows are seen on the same scale as the
radio jet (e.g. \citealt{Holt:2009}). Several recent integral field
spectroscopic (IFS) observations of luminous (radio-quiet) Type II AGN
(e.g. \citealt{Liu:2013, Harrison:2014, McElroy:2015}) have confirmed
the prevalence of large-scale AGN-driven winds. A similar spatially
resolved IFS study of the narrow line emission would confirm the
origin of the ionized gas kinematics in PKS\,B1740$-$517.

\subsection{Archival data from \emph{XMM-Newton}}

\subsubsection{Data analysis}

\begin{figure}
  \centering
  \includegraphics[width=0.5\textwidth, angle=0, trim=0 0 0 0]{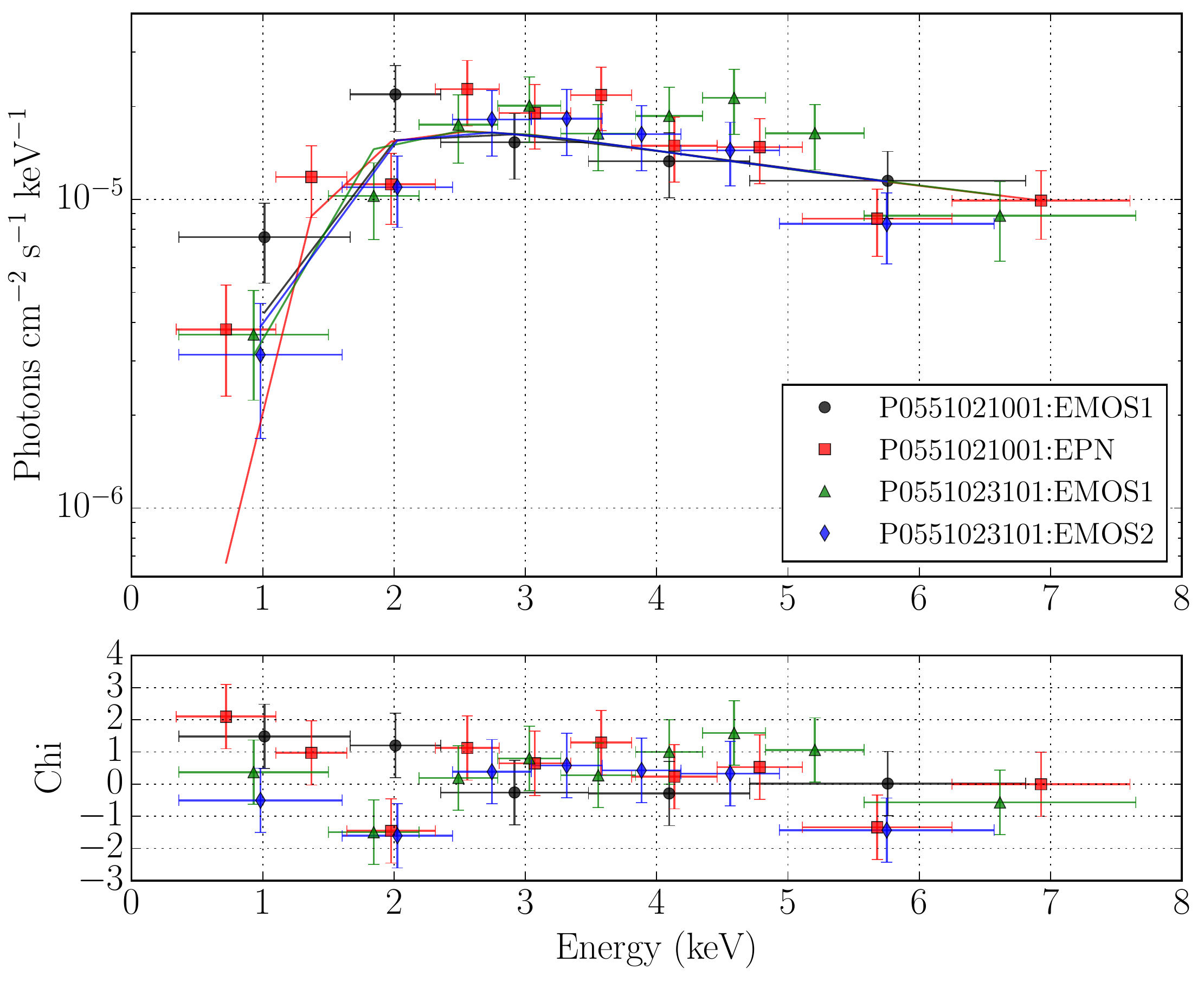}
  \caption{Upper panel: four X-ray spectra towards PKS\,B1740$-$517,
    observed with the \emph{XMM-Newton} satellite. The solid lines
    represent the best-fitting model produced using an absorbed
    power-law in \texttt{XSPEC}. Lower panel: the residuals from the
    best-fitting models, indicating scatter consistent with the
    noise.}
  \label{figure:PKS1740-517_xray_spectrum}
\end{figure}

By matching the ICRF2 position of PKS\,B1740$-$517 to sources listed
in the 3XMM-DR4 \emph{XMM-Newton} Serendipitous Source Catalogue, we
found a bright X-ray source, 3XMM\,J174425.3$-$514444, at an angular
separation of 1\,arcsec. This field was serendipitously observed as
part of targeted observations towards HD\,160691, a fifth-magnitude
star also known as $\mu$\,Arae, which is 5.9\,arcmin away from
PKS\,B1740$-$517. Via the \emph{XMM-Newton} Science
Archive\footnote{\url{http://xmm.esac.esa.int/xsa/}}, we obtained four
X-ray spectra of PKS\,B1740$-$517 (see
Fig.\,\ref{figure:PKS1740-517_xray_spectrum}), from two separate
observations of integration time 7800 and 11\,700\,s, and fitted these
simultaneously using the modelling package \texttt{XSPEC}
\citep{Arnaud:1996}. We found that the X-ray spectra are well fitted
by a standard absorbed power-law model. Because of the relatively low
Galactic latitude of $b = -11\fdg5$, we incorporate an additional
fixed absorption component in our model due to the Galactic
foreground, estimated within \texttt{XSPEC} to be $N_{\rm H,Gal} =
1.09\,\times\,10^{21}\,\mathrm{cm}^{-2}$. Using this model, we obtain
an estimate of the intrinsic column density towards PKS\,B1740$-$517
of $N_{\rm H,X} =
1.21^{+0.61}_{-0.43}\,\times\,10^{22}\,\mathrm{cm}^{-2}$, a photon
index of $\Gamma = 0.80^{+0.28}_{-0.24}$ and a 1\,keV normalization of
$4.8^{+2.8}_{-1.5} \times
10^{-5}$\,photons\,cm$^{-2}$\,s$^{-1}$\,keV$^{-1}$. While the
intrinsic column density is consistent with that measured for other
X-ray observed GPS radio sources, the photon index is
harder\footnote{$\langle{\Gamma}\rangle = 1.66$, $\sigma_{\Gamma} =
  0.36$ for the sample of 16 X-ray observed GPS sources discussed by
  \citet{Tengstrand:2009}.}, possibly indicating an obscured
Compton-thick AGN, similar to that seen for PKS\,B1607+26
(\citealt{Tengstrand:2009}). If the observed spectrum is indeed
arising from reflection of the primary X-ray emission off a
Compton-thick obscuring medium, then sight-lines to the AGN may
instead have column densities in excess of $10^{24}$\,cm$^{-2}$. Using
the above best fitting model to the \emph{XMM-Newton} spectra we
estimate an X-ray luminosity of $L_{2-10\,\mathrm{keV}} \approx
4.3\,\times 10^{43}$\,erg\,s$^{-1}$, consistent with the distribution
of luminosities measured for other GPS radio sources
(\citealt{Tengstrand:2009}).

\subsubsection{An X-ray upper limit on the HI spin
  temperature}\label{section:x-ray_spin_temp}

Combining the X-ray derived total column density of elemental hydrogen
and the 21-cm opacity data allows us to estimate an upper limit on the
\mbox{H\,{\sc i}} spin temperature, and compare this with direct
measurements of the spin temperature in other galaxies. In the absence
of any \mbox{H\,{\sc i}} emission, the 21-cm optical depth is given by
\begin{equation}\label{equation:optical_depth}
  \tau_{21} = -\ln{\left(1 + {\Delta{S}\over c_{\rm f}\,S_{\rm cont}}\right)},
\end{equation}
where $\Delta{S}$ is the change in flux density due to absorption,
$S_{\rm cont}$ is the continuum flux density and $c_{\rm f}$ is the
covering factor, i.e. the fraction of background source obscured by
the foreground absorber. The column density of \mbox{H\,{\sc i}} gas
(in units of atoms\,cm$^{-2}$) is then given by
\begin{equation}\label{equation:column_density}
  N_{\rm HI} = 1.823\times10^{18}\,T_{\rm spin}\,\int{\tau_{21}(v)\mathrm{d}v},
\end{equation}
where $T_{\rm spin}$ is the 21-cm spin temperature (in units of K) and
$\int{\tau_{21}(v)\mathrm{d}v}$ is the rest-frame velocity integrated
optical depth (in units of km\,s$^{-1}$). For sight lines where
several gas components may exist at different spin temperatures,
$T_{\rm spin}$ is the column-density-weighted harmonic mean over those
components. Assuming that the 21-cm and X-ray spectra probe similar
sight lines through the gas, an upper limit on the spin temperature
can be obtained from
\begin{equation}\label{equation:spin_temperature}
  T_{\rm spin} \lesssim 5485 \left[{N_{\rm H,X}\over10^{22}\,\mathrm{cm}^{-2}}\right] \left[{\int{\tau_{21}(v)\mathrm{d}v}\over\mathrm{1\,km\,s}^{-1}}\right]^{-1} \mathrm{K},
\end{equation}
where $N_{\rm H,X}$ is the total column density of elemental hydrogen
(in atomic, ionized and molecular form) estimated from the
\emph{XMM-Newton} spectra.

We estimate the 21-cm optical depth from our BETA data using
Equation\,\ref{equation:optical_depth}, which of course depends on the
value we adopt for the covering factor ($c_{\rm f})$. Assuming a
covering factor of unity (which gives a lower limit to the optical
depth and hence a spin temperature upper limit), our best estimate of
the total integrated optical depth from the average BETA spectrum is
$2.73\pm0.10$\,km\,s$^{-1}$. Based on the column density of total
hydrogen derived from a simple absorbed power-law model of the X-ray
spectra, we obtain an upper limit on the spin temperature of $T_{\rm
  spin} \lesssim 2430\pm1050$\,K. The large fractional uncertainty in
the X-ray derived $N_{\rm H,X}$ results in a poor constraint on this
upper limit, yet is consistent with other direct measurements of
extragalactic spin temperatures, both from 21-cm observations of known
damped Ly$\alpha$ absorption systems towards quasars
(\citealt{Curran:2012b, Kanekar:2014}) and by combining the intrinsic
\mbox{H\,{\sc i}} absorption in quasars with extinction estimates from
their optical continuum (e.g. PKS\,B1549$-$790; \citealt{Holt:2006}).
While this agreement is reassuring, the assumption of similar sight
lines to the radio and X-ray sources is tentative at best, depending
on the extent of the radio source and mechanism generating the X-ray
emission.

\section{Discussion}\label{section:discussion}

The \mbox{H\,{\sc i}} absorption line profiles associated with radio
galaxies are typically complex, encoding small-scale variation in both
the 21-cm opacity and the background source structure that are often
difficult to disentangle and accurately model. However, we can attempt
to overcome this by amalgamating information from the integrated
\mbox{H\,{\sc i}} spectrum, the radio continuum and any available
multiwavelength data. In the following we discuss the inferred
properties of the background radio source, followed by the individual
components seen in the absorption spectrum and their most likely
physical interpretation.

\subsection{Properties of the background radio source}

PKS\,B1740$-$517 is a powerful young GPS radio source, which through
optical imaging and spectroscopy we have established is hosted by a
luminous Seyfert galaxy at $z = 0.44$. A least-squares fit to the
spectral energy distribution (SED) shown in
Fig.\,\ref{figure:PKS1740-517_radio_sed} yields a spectral age of
approximately $t_{\rm age} \approx 2500$\,yr, where we have assumed an
equipartition magnetic field (\citealt{Murgia:2003}). From VLBI
observations, King found that the 2.3-GHz flux density is contained
within two 10 mas-scale radio components, in a ratio of approximately
of 3.9 to 1 \citep{King:1994}. The angular separation of the two
components is 52\,mas, corresponding to a physical scale of $\sim
300$\,pc. From the integrated SED alone we cannot confirm whether the
components represent hot-spots associated with two jets or a core and
jet. In previous VLBI studies of compact sources
(\citealt{Stanghellini:1997,Snellen:2000,deVries:2009}) it was found
that GPS radio sources associated with galaxies (as opposed to
quasars) tend to exhibit compact double (CD) or compact symmetric
object (CSO) morphologies. In the following discussion we therefore
assume that the VLBI-scale structure seen in this source represents
radio emission associated with jet hotspots, but cannot rule out the
possibility of a core and jet without further spectral information for
the two source components.

The brightness asymmetry of the two jet components could be either
caused by an intrinsic difference in the radio luminosity, possibly
resulting from an inhomogeneous circumnuclear medium, or by an
orientation effect. In the latter case we can estimate the angle $i$
at which the radio jet axis is inclined to the line-of-sight, as
follows
\begin{equation}
  \mathrm{cos}(i) = \left[{\gamma^{2}\over \gamma^{2}-1}\right]^{1/2}~\left[{R_{\rm jc}^{1/(2-\alpha)}-1\over R_{\rm jc}^{1/(2-\alpha)}+1}\right],
\end{equation}
where $R_{\rm jc}$ is the brightness ratio, $\alpha$ is the spectral
index and $\gamma$ is the bulk Lorentz factor in the jet. Assuming
that the intrinsic luminosities of the two jet components are equal
and that $\alpha = -0.74$, $R_{\rm jc} = 3.9$ (and a typical Lorentz
factor in the range $5 < \gamma < 10$), we estimate that the jet axis
is inclined by $i \approx 76\degr$ with respect to the line-of-sight,
giving a jet axis radius of $r \sim 150$\,pc. Based on the spectral
age we estimate an average separation velocity of approximately
0.4\,$c$ over the lifetime of the radio AGN, consistent with directly
measured expansion velocities of similarly powerful GPS and CSS
sources (see \citealt{deVries:2009} and references therein).  The
combined linear size, X-ray luminosity and 5-GHz radio
luminosity\footnote{For comparison with the literature we define the
  5-GHz luminosity as $L_{5\,{\rm GHz}} \equiv \nu_{5\,{\rm
      GHz}}\,S_{5\,{\rm GHz}}$, where $\nu_{5\,{\rm GHz}}$ is the
  rest-frame frequency and $S_{5\,{\rm GHz}}$ is the luminosity
  density.} ($L_{5\,{\rm GHz}} \approx 8.7 \times
10^{43}$\,erg\,s$^{-1}$) are all typical of other GPS radio sources in
the literature (e.g. \citealt{Tengstrand:2009}).

\subsection{The cold neutral gas in PKS\,B1740$-$517}

\subsubsection{Absorption components 1 and 2}

Absorption components 1 and 2 in the BETA spectrum have similar
velocities and may either be separate cold clouds of \mbox{H\,{\sc i}}
or the same structure seen against both source components. With a
width of $\Delta{v_{50}} \lesssim 5$\,km\,s$^{-1}$, the deep component
is unusually narrow for extragalactic \mbox{H\,{\sc i}} absorption
associated with the host galaxies of radio
AGN\footnote{\citet{Borthakur:2014} also found very narrow intervening
  absorption through the dwarf galaxy UCG\,7408 ($\Delta{v_{50}} =
  1.1$\,km\,s$^{-1}$), corresponding to cold gas at a kinetic
  temperature of $T_{\rm k} \approx 26$\,K.}. Typical widths often
range from more than 10 to 1000\,km\,s$^{-1}$ and are largely
attributed to turbulent and bulk motion of gas that is either rotating
or radially outflowing/infalling with respect to the nucleus (see
e.g. \citealt{Allison:2013}, \citealt{Gereb:2015} and references
therein). Narrow absorption is often found near to the systemic
velocity and thought to be associated with gas in a regularly rotating
structure, such as a disc or ring, through which a single sight line
to a compact radio source picks out the velocity dispersion.

In Fig.\,\ref{figure:PKS1740-517_line_comparison} we compare the
velocity structure of the \mbox{H\,{\sc i}} absorption and optical
emission lines. The prominent narrow component is blue shifted by
approximately $-200 \pm 50$\,km\,s$^{-1}$ with respect to the systemic
velocity and also the dynamical centre of the \mbox{[O\,{\sc iii}]}
emission lines. Assuming that the radio source is co-spatial with the
AGN, this offset would indicate that the gas is either rotating or
outflowing radially towards the observer from the nucleus. A similar
200\,km\,s$^{-1}$ velocity offset is seen for narrow \mbox{H\,{\sc i}}
absorption ($\Delta{v_{50}} \sim 90$\,km\,s$^{-1}$) in the nearby
Seyfert 2 galaxy NGC\,2110.  Detailed modelling by
\cite{Gallimore:1999} showed that the absorbing gas is located in a
$\sim 200$-pc disc, close to the radio source and with similar
orientation to the jet axis. The velocity offset seen here would also
suggest that the separation of the gas and radio source is small,
indicating a location that is $\sim 150$\,pc away from the central
nucleus. Since a jet-driven radial outflow would likely result in
significant turbulence of the gas, and therefore the narrow absorption
would represent an intact cloud that has become entrained, it is more
likely that we are seeing absorption through a foreground structure.

It is also possible that an unforeseen systematic error in the Gemini
spectrum (e.g. contamination from dust obscuration) may have caused an
artificial offset between the optical redshift and the narrow
absorption. If the narrow feature is instead at the systemic velocity,
this would place the absorbing gas at a distance much further out in
the galaxy (similar to PKS\,1814$-$637; \citealt{Morganti:2011}), and
would represent tangentially rotating gas seen in absorption against
the central radio AGN. Similarity between the line ratios of
absorption components 1 and 2 and the flux densities of the radio
components would also suggest lines of sight to each. Under this
scenario the distance of the absorbing gas from the nucleus would
simply be given by
\begin{equation}
  d_{\rm gas} \approx r_{\rm source} {v_{\rm disc}\over\Delta{v}_{\rm abs}} \sim 4\,\mathrm{kpc},
\end{equation}
where $r_{\rm source}$ is the physical separation of the two radio
components, $v_{\rm disc}$ is the rotational velocity of the gas
(assumed to be $\sim 200$\,km\,s$^{-1}$) and $\Delta{v}_{\rm abs}$ is
the velocity separation of the two absorption components. However,
there is no obvious evidence of the effect of dust obscuration in the
Gemini data and verifying this would require deeper optical imaging
and resolved spectroscopy.

The peak \mbox{H\,{\sc i}} component is sufficiently narrow that we
can estimate a reasonable upper limit on the kinetic temperature of
the gas. We use the following relationship between the temperature and
line width
\begin{equation}\label{equation:kinetic_temperature}
  T_{\rm k} \lesssim {1 \over 8\ln(2)} {m_{\rm H} \over k_{\rm B}} {(\Delta{v}_{50})^{2}} = 538 \pm 35\,\mathrm{K},
\end{equation}
where $\Delta{v}_{50}$ is the FWHM of the spectral line, $m_{\rm H}$
is the mass of a hydrogen atom and $k_{\rm B}$ is the Boltzmann
constant. Given that turbulent motion of the gas will result in line
broadening, we expect that the true kinetic temperature could be
significantly less than this and in a phase similar to the CNM of the
Milky Way ($T_{\rm k} \sim 40$--$200$\,K;
\citealt{Wolfire:2003}). Assuming that the gas is dense enough to be
collisionally excited, the kinetic temperature is a reasonable proxy
for the \mbox{H\,{\sc i}} spin temperature, and provides a stronger
constraint than the X-ray upper limit derived in
Section\,\ref{section:x-ray_spin_temp}. However, irradiation by UV
emission from the AGN and continuum emission from the powerful radio
source may drive the spin temperature to much higher values
(\citealt{Bahcall:1969}). Given the high luminosity of the radio
source, and the fact that we detect \mbox{H\,{\sc i}} absorption, we
find it improbable that the spin temperature could be dominated by
21-cm radiation. We therefore assume that the gas is sufficiently
dense and/or shielded from the continuum to be collisionally excited,
but acknowledge that otherwise the spin temperature could be an order
of magnitude higher than the kinetic temperature.

If the peak absorption is located in front of the fainter radio
component it would imply that all of the radio flux is being absorbed
(given the flux ratio of 3.9 to 1 at 2.3\,GHz), in which case we would
expect the profile to be saturated. Given that the ASKAP spectral
channels do not fully sample this narrow line\footnote{Making this
  system an excellent future target for proposed high spectral
  resolution zoom modes with the SKA pathfinders.}, we cannot
adequately resolve the profile to test for saturation. However,
resolved \mbox{H\,{\sc i}} absorption lines associated with radio AGN
do not typically exhibit saturation and such a scenario would indicate
that the gas here is considerably colder than the upper limit given in
Equation\,\ref{equation:kinetic_temperature}. We therefore favour the
scenario that the peak absorption is seen in front of the brighter
radio source. In which case the \mbox{H\,{\sc i}} covering factor of
the total radio flux is $c_{\rm f} = 0.8$, and so the velocity
integrated optical depth for the deep component is
$1.55\pm0.07$\,km\,s$^{-1}$. Using the above upper limit to the
kinetic temperature, and the relationship given by
Equation\,\ref{equation:column_density}, we estimate a corresponding
column density of $N_{\rm HI} \lesssim (1.52\pm0.12) \times
10^{21}$\,cm$^{-2}$. The angular size of the brighter radio component
measured by \cite{King:1994} is $6.7$\,mas $\times 5.8$\,mas (at
2.3\,GHz), which at a redshift of $z = 0.44$ equates to a physical
scale of $38 \times 33$\,pc$^{2}$. We therefore estimate a mass of
$M_{\rm HI} \sim 5\times 10^{4}\,\mathrm{M}_{\odot}$ for the
foreground gas cloud. Uncertainties in the spin temperature and
covering factor mean that the column density and mass could be an
order of magnitude larger.

\subsubsection{Absorption components 3 and 4}

Component 3 (Fig.\,\ref{figure:PKS1740-517_beta_fit}) is consistent
with the dynamical centre of the \mbox{[O\,{\sc iii}]} emission and
offset by approximately $-100 \pm 50$\,km\,s$^{-1}$ with respect to
the systemic velocity. Its width ($\Delta{v}_{50} \approx
50$\,km\,s$^{-1}$) is comparable with \mbox{H\,{\sc i}} absorption
often seen through the inclined discs of nearby Seyfert galaxies
(e.g. \citealt{Dickey:1982, Gallimore:1999, Allison:2014}). It is
likely that component 3 is arising from tangentially rotating
\mbox{H\,{\sc i}} gas in a disc or ring, which would give a radial
velocity close to that of the system. On the other side of the
absorption complex we tentatively see a weak, broad 300\,km\,s$^{-1}$
feature (component 4) shifted by $-350 \pm 50$\,km\,s$^{-1}$ with
respect to the system. Such a profile is typical of jet-driven neutral
outflows seen in the host galaxies of other powerful radio AGN and we
discuss this further in Section\,\ref{section:outflow}.

Under the alternative scenario discussed above, whereby the deep
narrow absorption is actually at the systemic redshift (and an unknown
systematic error is causing a velocity offset), we infer a similar
model to that of PKS\,B1814$-$637 (\citealt{Morganti:2011}), where the
narrow absorption is arising from gas in the larger galaxy and the
broader features are the result of a rotating circumnuclear disc close
to the radio source. However, there is no evidence so far to suggest
that the optical redshift is being corrupted by a systematic error,
and further observations at higher spatial and spectral resolution
would shed light on this possibility.

\subsection{A possible jet-driven neutral gas
  outflow}\label{section:outflow}

Jet-driven neutral gas outflows are direct indicators of the feedback
between a powerful radio AGN and the surrounding ISM
(\citealt{Hardcastle:2007}). Gas outflows are found to be ubiquitous
in the host galaxies of powerful compact radio AGN, and signify the
growth of the source as it expands through a natal cocoon of dust and
gas (e.g. \citealt{Holt:2009}). In the case of PKS\,B1740$-$517,
absorption component 4 (Fig.\,\ref{figure:PKS1740-517_beta_fit}) may
represent a neutral gas outflow that is being driven by the
approaching jet as it drives into the circumnuclear gas. The mass
outflow rate of the neutral gas can be estimated by assuming a simple
superwind model (\citealt{Heckman:2000}):
\begin{equation}
  \dot{M}_{\rm HI} = 30{\Omega\over 4\mathrm{\pi}}{r\over 1\,\mathrm{kpc}}{N_{\rm HI}\over 10^{21}\,\mathrm{cm}^{-2}}{v\over 300\,\mathrm{km}\,\mathrm{s}^{-1}}\,\mathrm{M}_{\odot}\,\mathrm{yr}^{-1},
\end{equation}
where $\Omega$ is the solid angle into which the gas is flowing
(assumed to be $\sim\mathrm{\pi}$), $r$ is the radius, $v$ is the
velocity of the outflow, and $N_{\rm HI}$ is the column density of
\mbox{H\,{\sc i}} gas.  Using an outflow velocity of $v \sim
300$\,km\,s$^{-1}$ and a column density estimate of $N_{\rm HI} \sim 1
\times 10^{21}$\,cm$^{-2}$, which assumes that the outflowing gas has
a similar spin temperature to that of the narrow absorption, we
estimate an outflow rate of
$\sim$1\,$\mathrm{M}_{\odot}$\,yr$^{-1}$. This is strongly dependent
upon our assumptions about the geometry and spin temperature of the
gas, and we note that the outflow rate may be an order of magnitude
higher. Within this framework the narrow absorption may either
represent unperturbed circumnuclear gas or a dense cloud that has
become entrained in the outflow.

\section{Conclusion and future prospects}

Using commissioning data from the six-antenna BETA of ASKAP we have
discovered \mbox{H\,{\sc i}} 21-cm absorption that is intrinsic to the
host galaxy of the powerful compact radio source PKS\,B1740$-$517. Our
result demonstrates the excellent potential of ASKAP to search for
\mbox{H\,{\sc i}} in a continuous redshift range of $z =$ 0.4 -- 1.0,
equivalent to look back times in the range 4.2 -- 7.7\,Gyr. Such a
capability is made possible by the excellent radio quiet environment
of the observatory site at these frequencies. Furthermore, the
detection and characterization of a complex absorption profile is very
encouraging for the study of the neutral environments of radio AGN in
future wide-field, broad-band absorption surveys with ASKAP and other
SKA pathfinders.

Optical spectroscopy with the 8-m Gemini-South telescope confirms that
the \mbox{H\,{\sc i}} absorption is intrinsic to the host galaxy of
the radio source. The galaxy has a luminous active nucleus, with
several forbidden oxygen lines exhibiting double-peaked structure on
large-scales that indicate possible AGN-driven outflows of ionized
gas. Archival data from the \emph{XMM-Newton} satellite show that the
soft X-ray emission is absorbed by a dense obscuring medium, possibly
Compton thick. The profile of the \mbox{H\,{\sc i}} line is complex
and exhibits the typical narrow and broad features associated with
absorption in the host galaxy of a compact radio AGN.  Based on the
significantly blueshifted broad absorption we conclude that
PKS\,B1740$-$517 represents a recently triggered radio source ($t_{\rm
  age} \approx 2500$\,yr) in which the jets are breaking through the
surrounding cocoon of dense circumnuclear gas. This object requires
further study, with spatially resolved optical and 21-cm spectroscopy,
and deeper optical imaging, to fully understand the neutral and
ionized gas kinematics.

\section*{Acknowledgements} 

We thank Sean Farrell, Davide Burlon, Rebecca McElroy and David
Parkinson for useful discussions. We also thank the anonymous referee
and Flornes Yuen for useful comments that help to improve this paper.

The Australian SKA Pathfinder is part of the Australia Telescope
National Facility which is managed by CSIRO. Operation of ASKAP is
funded by the Australian Government with support from the National
Collaborative Research Infrastructure Strategy. Establishment of the
Murchison Radio-astronomy Observatory was funded by the Australian
Government and the Government of Western Australia.  ASKAP uses
advanced supercomputing resources at the Pawsey Supercomputing
Centre. We acknowledge the Wajarri Yamatji people as the traditional
owners of the Observatory site.

Supporting observations were obtained at the Gemini Observatory, which
is operated by the Association of Universities for Research in
Astronomy, Inc., under a cooperative agreement with the NSF on behalf
of the Gemini partnership: the National Science Foundation (USA), the
National Research Council (Canada), CONICYT (Chile), the Australian
Research Council (Australia), Minist\'{e}rio da Ci\^{e}ncia,
Tecnologia e Inova\c{c}\~{a}o (Brazil) and Ministerio de Ciencia,
Tecnolog\'{i}a e Innovaci\'{o}n Productiva (Argentina). X-ray data
were based on observations obtained with \emph{XMM-Newton}, an ESA
science mission with instruments and contributions directly funded by
ESA member states and NASA.

JRA acknowledges support from a Bolton Fellowship. RM gratefully
acknowledges support from the European Research Council under the
European Union's Seventh Framework Programme (FP/2007-2013) ERC
Advanced Grant RADIOLIFE-320745.  Parts of this research were
conducted by the Australian Research Council Centre of Excellence for
All-sky Astrophysics (CAASTRO), through project number CE110001020. We
have made use of \texttt{Astropy}, a community-developed core
\texttt{Python} package for astronomy (Astropy Collaboration, 2013);
the NASA/IPAC Extragalactic Database (NED) which is operated by the
Jet Propulsion Laboratory, California Institute of Technology, under
contract with the National Aeronautics and Space Administration;
NASA's Astrophysics Data System Bibliographic Services; the SIMBAD
data base and VizieR catalogue access tool, both operated at CDS,
Strasbourg, France.

The authors would also like to acknowledge the large group of people
who have contributed to the planning, design, construction and support
of BETA and ASKAP. This includes: Kerry Ardern, Brett Armstrong, Jay
Banyer, Samantha Barry, Tim Bateman, Ron Beresford, Brayden Briggs,
Kate Brooks, Graeme Carrad, Ettore Carretti, Frank Ceccato, Raji
Chekkala, Kate Chow, Geoff Cook, Paul Cooper, Evan Davis, Ludovico de
Souza, Jack Dixon, Peter Duffy, Troy Elton, Brian Jeffs, Alex Harding,
George Hobbs, Ron Koenig, Arkadi Kosmynin, Tom Lees, Amy Loke, Li Li,
Stacy Mader, Tony Maher, Neil Marston, Vincent McIntyre, Ian McRobert,
Samantha Mickle, Ray Moncay, Neale Morison, John Morris, Tony Mulry,
Alan Ng, Wilfredo Pena, Nathan Pope, Brett Preisig, Lou Puls, Michael
Reay, Ken Reeves, Adrian Rispler, Victor Rodrigues, Daniel Roxby, Tim
Ruckley, Craig Russell, Aaron Sanders, Ken Smart, Mark Wieringa, Tim
Wilson, Kjetil Wormnes and Xinyu Wu.

We would finally like to acknowledge the contributions of the ASKAP
survey science teams, represented by the following group leaders:
Shami Chatterjee, John Dickey, Bryan Gaensler, Peter Hall, Tom
Landecker, Martin Meyer, Tara Murphy, Ingrid Stairs, Lister
Staveley-Smith, Russ Taylor and Steven Tingay.

\bibliographystyle{mn2e.bst}
\bibliography{./bibliography.bib}

\begin{thebibliography}{135}
\expandafter\ifx\csname natexlab\endcsname\relax\def\natexlab#1{#1}\fi

\bibitem[{{Allison} {et~al}\mbox{.}(2012){Allison}, {Curran}, {Emonts},
  {Ger{\'e}b}, {Mahony}, {Reeves}, {Sadler}, {Tanna}, {Whiting}, \&
  {Zwaan}}]{Allison:2012a}
{Allison} J.~R. {et~al.}, 2012, MNRAS, 423, 2601

\bibitem[{{Allison} {et~al}\mbox{.}(2013){Allison}, {Curran}, {Sadler}, \&
  {Reeves}}]{Allison:2013}
{Allison} J.~R., {Curran} S.~J., {Sadler} E.~M., {Reeves} S.~N., 2013, MNRAS,
  430, 157

\bibitem[{{Allison} {et~al}\mbox{.}(2014){Allison}, {Sadler}, \&
  {Meekin}}]{Allison:2014}
{Allison} J.~R., {Sadler} E.~M., {Meekin} A.~M., 2014, MNRAS, 440, 696

\bibitem[{Arnaud(1996)}]{Arnaud:1996}
Arnaud K.~A., 1996, in ASP Conf. Ser., Vol. 101, Astronomical Data Analysis
  Software and Systems V, {Jacoby} G.~H., {Barnes} J., eds., Astron. Soc. Pac.,
  San Francisco, p.~17

\bibitem[{{Bahcall} \& {Ekers}(1969)}]{Bahcall:1969}
{Bahcall} J.~N., {Ekers} R.~D., 1969, ApJ, 157, 1055

\bibitem[{{Blain} {et~al}\mbox{.}(2013){Blain}, {Assef}, {Stern}, {Tsai},
  {Eisenhardt}, {Bridge}, {Benford}, {Jarrett}, {Cutri}, {Petty}, {Wu}, \&
  {Wright}}]{Blain:2013}
{Blain} A.~W. {et~al.}, 2013, ApJ, 778, 113 (erratum: ApJ, 782, 58)

\bibitem[{{Booth} {et~al}\mbox{.}(2009){Booth}, {de Blok}, {Jonas}, \&
  {Fanaroff}}]{Booth:2009}
{Booth} R.~S., {de Blok} W.~J.~G., {Jonas} J.~L., {Fanaroff} B., 2009,
  arXiv:0910.2935

\bibitem[{{Borthakur} {et~al}\mbox{.}(2014){Borthakur}, {Momjian}, {Heckman},
  {York}, {Bowen}, {Yun}, \& {Tripp}}]{Borthakur:2014}
{Borthakur} S., {Momjian} E., {Heckman} T.~M., {York} D.~G., {Bowen} D.~V.,
  {Yun} M.~S., {Tripp} T.~M., 2014, ApJ, 795, 98

\bibitem[{{Braun}(2012)}]{Braun:2012}
{Braun} R., 2012, ApJ, 749, 87

\bibitem[{{Briggs}(1983)}]{Briggs:1983}
{Briggs} F.~H., 1983, ApJ, 274, 86

\bibitem[{{Burgess} \& {Hunstead}(2006)}]{Burgess:2006}
{Burgess} A.~M., {Hunstead} R.~W., 2006, AJ, 131, 114

\bibitem[{{Campbell-Wilson} \& {Hunstead}(1994)}]{Campbell-Wilson:1994}
{Campbell-Wilson} D., {Hunstead} R.~W., 1994, Proc. of the Astro. Soc. Aust.,
  11, 33

\bibitem[{{Cappellari} \& {Emsellem}(2004)}]{Cappellari:2004}
{Cappellari} M., {Emsellem} E., 2004, PASP, 116, 138

\bibitem[{{Carilli} {et~al}\mbox{.}(1998){Carilli}, {Menten}, {Reid}, {Rupen},
  \& {Yun}}]{Carilli:1998}
{Carilli} C.~L., {Menten} K.~M., {Reid} M.~J., {Rupen} M.~P., {Yun} M.~S.,
  1998, ApJ, 494, 175

\bibitem[{{Catinella} {et~al}\mbox{.}(2008){Catinella}, {Haynes}, {Giovanelli},
  {Gardner}, \& {Connolly}}]{Catinella:2008}
{Catinella} B., {Haynes} M.~P., {Giovanelli} R., {Gardner} J.~P., {Connolly}
  A.~J., 2008, ApJ, 685, L13

\bibitem[{{Chandola} {et~al}\mbox{.}(2011){Chandola}, {Sirothia}, \&
  {Saikia}}]{Chandola:2011}
{Chandola} Y., {Sirothia} S.~K., {Saikia} D.~J., 2011, MNRAS, 418, 1787

\bibitem[{{Chang} {et~al}\mbox{.}(2010){Chang}, {Pen}, {Bandura}, \&
  {Peterson}}]{Chang:2010}
{Chang} T.-C., {Pen} U.-L., {Bandura} K., {Peterson} J.~B., 2010, {Nature},
  466, 463

\bibitem[{{Condon} {et~al}\mbox{.}(1998){Condon}, {Cotton}, {Greisen}, {Yin},
  {Perley}, {Taylor}, \& {Broderick}}]{Condon:1998}
{Condon} J.~J., {Cotton} W.~D., {Greisen} E.~W., {Yin} Q.~F., {Perley} R.~A.,
  {Taylor} G.~B., {Broderick} J.~J., 1998, AJ, 115, 1693

\bibitem[{{Curran}(2012)}]{Curran:2012b}
{Curran} S.~J., 2012, ApJ, 748, L18

\bibitem[{{Curran} {et~al}\mbox{.}(2013{\natexlab{a}}){Curran}, {Allison},
  {Glowacki}, {Whiting}, \& {Sadler}}]{Curran:2013b}
{Curran} S.~J., {Allison} J.~R., {Glowacki} M., {Whiting} M.~T., {Sadler}
  E.~M., 2013{\natexlab{a}}, MNRAS, 431, 3408

\bibitem[{{Curran} {et~al}\mbox{.}(2011){Curran}, {Whiting}, {Murphy}, {Webb},
  {Bignell}, {Polatidis}, {Wiklind}, {Francis}, \& {Langston}}]{Curran:2011a}
{Curran} S.~J. {et~al.}, 2011, MNRAS, 413, 1165

\bibitem[{{Curran} {et~al}\mbox{.}(2006){Curran}, {Whiting}, {Murphy}, {Webb},
  {Longmore}, {Pihlstr{\"o}m}, {Athreya}, \& {Blake}}]{Curran:2006a}
{Curran} S.~J., {Whiting} M.~T., {Murphy} M.~T., {Webb} J.~K., {Longmore}
  S.~N., {Pihlstr{\"o}m} Y.~M., {Athreya} R., {Blake} C., 2006, MNRAS, 371, 431

\bibitem[{{Curran} {et~al}\mbox{.}(2013{\natexlab{b}}){Curran}, {Whiting},
  {Sadler}, \& {Bignell}}]{Curran:2013a}
{Curran} S.~J., {Whiting} M.~T., {Sadler} E.~M., {Bignell} C.,
  2013{\natexlab{b}}, MNRAS, 428, 2053

\bibitem[{{Curran} {et~al}\mbox{.}(2008){Curran}, {Whiting}, {Wiklind}, {Webb},
  {Murphy}, \& {Purcell}}]{Curran:2008}
{Curran} S.~J., {Whiting} M.~T., {Wiklind} T., {Webb} J.~K., {Murphy} M.~T.,
  {Purcell} C.~R., 2008, MNRAS, 391, 765

\bibitem[{{de Vries} {et~al}\mbox{.}(2009){de Vries}, {Snellen}, {Schilizzi},
  {Mack}, \& {Kaiser}}]{deVries:2009}
{de Vries} N., {Snellen} I.~A.~G., {Schilizzi} R.~T., {Mack} K.-H., {Kaiser}
  C.~R., 2009, A\&A, 498, 641

\bibitem[{{de Zotti} {et~al}\mbox{.}(2010){de Zotti}, {Massardi}, {Negrello},
  \& {Wall}}]{deZotti:2010}
{de Zotti} G., {Massardi} M., {Negrello} M., {Wall} J., 2010, A\&AR, 18, 1

\bibitem[{{Deboer} {et~al}\mbox{.}(2009){Deboer}, {Gough}, {Bunton},
  {Cornwell}, {Beresford}, {Johnston}, {Feain}, {Schinckel}, {Jackson}, \&
  et~al.}]{Deboer:2009}
{Deboer} D.~R. {et~al.}, 2009, Proc. IEEE, 97, 1507

\bibitem[{{Delhaize} {et~al}\mbox{.}(2013){Delhaize}, {Meyer},
  {Staveley-Smith}, \& {Boyle}}]{Delhaize:2013}
{Delhaize} J., {Meyer} M.~J., {Staveley-Smith} L., {Boyle} B.~J., 2013, MNRAS,
  433, 1398

\bibitem[{{di Serego-Alighieri} {et~al}\mbox{.}(1994){di Serego-Alighieri},
  {Danziger}, {Morganti}, \& {Tadhunter}}]{DiSerego-Alighieri:1994}
{di Serego-Alighieri} S., {Danziger} I.~J., {Morganti} R., {Tadhunter} C.~N.,
  1994, MNRAS, 269, 998

\bibitem[{{Dickey}(1982)}]{Dickey:1982}
{Dickey} J.~M., 1982, ApJ, 263, 87

\bibitem[{{Dopita} \& {Sutherland}(1995)}]{Dopita:1995}
{Dopita} M.~A., {Sutherland} R.~S., 1995, ApJ, 455, 468

\bibitem[{{Emonts} {et~al}\mbox{.}(2010){Emonts}, {Morganti}, {Struve},
  {Oosterloo}, {van Moorsel}, {Tadhunter}, {van der Hulst}, {Brogt}, {Holt}, \&
  {Mirabal}}]{Emonts:2010}
{Emonts} B.~H.~C. {et~al.}, 2010, MNRAS, 406, 987

\bibitem[{{Fanti} {et~al}\mbox{.}(1995){Fanti}, {Fanti}, {Dallacasa},
  {Schilizzi}, {Spencer}, \& {Stanghellini}}]{Fanti:1995}
{Fanti} C., {Fanti} R., {Dallacasa} D., {Schilizzi} R.~T., {Spencer} R.~E.,
  {Stanghellini} C., 1995, A\&A, 302, 317

\bibitem[{{Fey} {et~al}\mbox{.}(2009){Fey}, {Gordon}, \& {Jacobs}}]{Fey:2009}
{Fey} A., {Gordon} G., {Jacobs} C., eds., 2009, {The Second Realization of the
  International Celestial Reference Frame by VLBI, IERS Technical Notes 35}.
  Verlad des Bundesamts fur Kartographie und Geodasie, Frankfurt am Main

\bibitem[{{Fischer} {et~al}\mbox{.}(2011){Fischer}, {Crenshaw}, {Kraemer},
  {Schmitt}, {Mushotsky}, \& {Dunn}}]{Fischer:2011}
{Fischer} T.~C., {Crenshaw} D.~M., {Kraemer} S.~B., {Schmitt} H.~R.,
  {Mushotsky} R.~F., {Dunn} J.~P., 2011, ApJ, 727, 71

\bibitem[{{Freudling} {et~al}\mbox{.}(2011){Freudling}, {Staveley-Smith},
  {Catinella}, {Minchin}, {Calabretta}, {Momjian}, {Zwaan}, {Meyer}, \&
  {O'Neil}}]{Freudling:2011}
{Freudling} W. {et~al.}, 2011, ApJ, 727, 40

\bibitem[{{Fu} {et~al}\mbox{.}(2012){Fu}, {Yan}, {Myers}, {Stockton},
  {Djorgovski}, {Aldering}, \& {Rich}}]{Fu:2012}
{Fu} H., {Yan} L., {Myers} A.~D., {Stockton} A., {Djorgovski} S.~G., {Aldering}
  G., {Rich} J.~A., 2012, ApJ, 745, 67

\bibitem[{{Gallimore} {et~al}\mbox{.}(1999){Gallimore}, {Baum}, {O'Dea},
  {Pedlar}, \& {Brinks}}]{Gallimore:1999}
{Gallimore} J.~F., {Baum} S.~A., {O'Dea} C.~P., {Pedlar} A., {Brinks} E., 1999,
  ApJ, 524, 684

\bibitem[{{Ger{\'e}b} {et~al}\mbox{.}(2015){Ger{\'e}b}, {Maccagni}, {Morganti},
  \& {Oosterloo}}]{Gereb:2015}
{Ger{\'e}b} K., {Maccagni} F.~M., {Morganti} R., {Oosterloo} T.~A., 2015, A\&A,
  575, A44

\bibitem[{{Ger{\'e}b} {et~al}\mbox{.}(2014){Ger{\'e}b}, {Morganti}, \&
  {Oosterloo}}]{Gereb:2014}
{Ger{\'e}b} K., {Morganti} R., {Oosterloo} T.~A., 2014, A\&A, 569, A35

\bibitem[{{Gregory} {et~al}\mbox{.}(1994){Gregory}, {Vavasour}, {Scott}, \&
  {Condon}}]{Gregory:1994}
{Gregory} P.~C., {Vavasour} J.~D., {Scott} W.~K., {Condon} J.~J., 1994, ApJS,
  90, 173

\bibitem[{{Gupta} {et~al}\mbox{.}(2006){Gupta}, {Salter}, {Saikia}, {Ghosh}, \&
  {Jeyakumar}}]{Gupta:2006a}
{Gupta} N., {Salter} C.~J., {Saikia} D.~J., {Ghosh} T., {Jeyakumar} S., 2006,
  MNRAS, 373, 972

\bibitem[{{Hambly} {et~al}\mbox{.}(2001){Hambly}, {MacGillivray}, {Read},
  {Tritton}, {Thomson}, {Kelly}, {Morgan}, {Smith}, {Driver}, {Williamson},
  {Parker}, {Hawkins}, {Williams}, \& {Lawrence}}]{Hambly:2001}
{Hambly} N.~C. {et~al.}, 2001, MNRAS, 326, 1279

\bibitem[{{Hardcastle} {et~al}\mbox{.}(2007){Hardcastle}, {Evans}, \&
  {Croston}}]{Hardcastle:2007}
{Hardcastle} M.~J., {Evans} D.~A., {Croston} J.~H., 2007, MNRAS, 376, 1849

\bibitem[{{Harrison} {et~al}\mbox{.}(2014){Harrison}, {Alexander}, {Mullaney},
  \& {Swinbank}}]{Harrison:2014}
{Harrison} C.~M., {Alexander} D.~M., {Mullaney} J.~R., {Swinbank} A.~M., 2014,
  MNRAS, 441, 3306

\bibitem[{{Hay} \& {O'Sullivan}(2008)}]{Hay:2008}
{Hay} S.~G., {O'Sullivan} J.~D., 2008, Radio Sci., 43, 6

\bibitem[{{Haynes} {et~al}\mbox{.}(2011){Haynes}, {Giovanelli}, {Martin},
  {Hess}, {Saintonge}, {Adams}, {Hallenbeck}, {Hoffman}, {Huang}, {Kent},
  {Koopmann}, {Papastergis}, {Stierwalt}, {Balonek}, {Craig}, {Higdon},
  {Kornreich}, {Miller}, {O'Donoghue}, {Olowin}, {Rosenberg}, {Spekkens},
  {Troischt}, \& {Wilcots}}]{Haynes:2011}
{Haynes} M.~P. {et~al.}, 2011, AJ, 142, 170

\bibitem[{{Heald} {et~al}\mbox{.}(2011){Heald}, {J{\'o}zsa}, {Serra},
  {Zschaechner}, {Rand}, {Fraternali}, {Oosterloo}, {Walterbos}, {J{\"u}tte},
  \& {Gentile}}]{Heald:2011}
{Heald} G. {et~al.}, 2011, A\&A, 526, A118

\bibitem[{{Healey} {et~al}\mbox{.}(2007){Healey}, {Romani}, {Taylor}, {Sadler},
  {Ricci}, {Murphy}, {Ulvestad}, \& {Winn}}]{Healey:2007}
{Healey} S.~E., {Romani} R.~W., {Taylor} G.~B., {Sadler} E.~M., {Ricci} R.,
  {Murphy} T., {Ulvestad} J.~S., {Winn} J.~N., 2007, ApJS, 171, 61

\bibitem[{{Heckman}(1980)}]{Heckman:1980}
{Heckman} T.~M., 1980, A\&A, 87, 152

\bibitem[{{Heckman} {et~al}\mbox{.}(2000){Heckman}, {Lehnert}, {Strickland}, \&
  {Armus}}]{Heckman:2000}
{Heckman} T.~M., {Lehnert} M.~D., {Strickland} D.~K., {Armus} L., 2000, ApJS,
  129, 493

\bibitem[{{H{\"o}gbom}(1974)}]{Hogbom:1974}
{H{\"o}gbom} J.~A., 1974, A\&AS, 15, 417

\bibitem[{{Holt} {et~al}\mbox{.}(2006){Holt}, {Tadhunter}, {Morganti},
  {Bellamy}, {Gonz{\'a}lez Delgado}, {Tzioumis}, \& {Inskip}}]{Holt:2006}
{Holt} J., {Tadhunter} C., {Morganti} R., {Bellamy} M., {Gonz{\'a}lez Delgado}
  R.~M., {Tzioumis} A., {Inskip} K.~J., 2006, MNRAS, 370, 1633

\bibitem[{{Holt} {et~al}\mbox{.}(2009){Holt}, {Tadhunter}, \&
  {Morganti}}]{Holt:2009}
{Holt} J., {Tadhunter} C.~N., {Morganti} R., 2009, MNRAS, 400, 589

\bibitem[{{Hook} {et~al}\mbox{.}(2003){Hook}, {Allington-Smith}, {Beard},
  {Crampton}, {Davies}, {Dickson}, {Ebbers}, {Fletcher}, {Jorgensen}, {Jean},
  {Juneau}, {Murowinski}, {Nolan}, {Laidlaw}, {Leckie}, {Marshall}, {Purkins},
  {Richardson}, {Roberts}, {Simons}, {Smith}, {Stilburn}, {Szeto}, {Tierney},
  {Wolff}, \& {Wooff}}]{Hook:2003}
{Hook} I. {et~al.}, 2003, in Proc. SPIE, Vol. 4841, Instrument Design and
  Performance for Optical/Infrared Ground Based Telescopes, {Iye} M.,
  {Moorwood} A.~F.~M., eds., SPIE, Bellingham, p. 1645

\bibitem[{Hotan {et~al}\mbox{.}(2014)Hotan, Bunton, Harvey-Smith, Humphreys,
  Jeffs, Shimwell, Tuthill, Voronkov, Allen, Amy, Ardern, Axtens, Ball,
  Bannister, Barker, Bateman, Beresford, Bock, Bolton, Bowen, Boyle, et~al.,
  Braun, \& et~al.}]{Hotan:2014}
Hotan A.~W. {et~al.}, 2014, Publ. Astron. Soc. Aust., 31, e041

\bibitem[{{Jacobs} {et~al}\mbox{.}(2011){Jacobs}, {Aguirre}, {Parsons},
  {Pober}, {Bradley}, {Carilli}, {Gugliucci}, {Manley}, {van der Merwe},
  {Moore}, \& {Parashare}}]{Jacobs:2011}
{Jacobs} D.~C. {et~al.}, 2011, ApJL, 734, L34

\bibitem[{{Jauncey} {et~al}\mbox{.}(2003){Jauncey}, {King}, {Bignall},
  {Lovell}, {Kedziora-Chudczer}, {Tzioumis}, {Tingay}, {Macquart}, \&
  {McCulloch}}]{Jauncey:2003}
{Jauncey} D.~L. {et~al.}, 2003, Publ. Astron. Soc. Aust., 20, 151

\bibitem[{{Johnston} {et~al}\mbox{.}(2007){Johnston}, {Bailes}, {Bartel},
  {Baugh}, {Bietenholz}, {Blake}, {Braun}, {Brown}, {Chatterjee}, {Darling},
  {Deller}, \& et~al.}]{Johnston:2007}
{Johnston} S. {et~al.}, 2007, Publ. Astron. Soc. Aust., 24, 174

\bibitem[{{Kanekar} {et~al}\mbox{.}(2011){Kanekar}, {Braun}, \&
  {Roy}}]{Kanekar:2011}
{Kanekar} N., {Braun} R., {Roy} N., 2011, ApJL, 737, L33

\bibitem[{{Kanekar} \& {Chengalur}(2001)}]{Kanekar:2001}
{Kanekar} N., {Chengalur} J.~N., 2001, MNRAS, 325, 631

\bibitem[{{Kanekar} \& {Chengalur}(2008)}]{Kanekar:2008}
{Kanekar} N., {Chengalur} J.~N., 2008, MNRAS, 384, L6

\bibitem[{{Kanekar} {et~al}\mbox{.}(2009){Kanekar}, {Prochaska}, {Ellison}, \&
  {Chengalur}}]{Kanekar:2009a}
{Kanekar} N., {Prochaska} J.~X., {Ellison} S.~L., {Chengalur} J.~N., 2009,
  MNRAS, 396, 385

\bibitem[{{Kanekar} {et~al}\mbox{.}(2014){Kanekar}, {Prochaska}, {Smette},
  {Ellison}, {Ryan-Weber}, {Momjian}, {Briggs}, {Lane}, {Chengalur},
  {Delafosse}, {Grave}, {Jacobsen}, \& {de Bruyn}}]{Kanekar:2014}
{Kanekar} N. {et~al.}, 2014, MNRAS, 438, 2131

\bibitem[{{King}(1994)}]{King:1994}
{King} E., 1994, PhD thesis, Univ. Tasmania, Hobart

\bibitem[{{Koribalski}(2010)}]{Koribalski:2010}
{Koribalski} B.~S., 2010, in ASP Conf. Ser., Vol. 421, Galaxies in Isolation:
  Exploring Nature Versus Nurture, {Verdes-Montenegro} L., {Del Olmo} A.,
  {Sulentic} J., eds., Astron. Soc. Pac., San Francisco, p. 137

\bibitem[{{Lagos} {et~al}\mbox{.}(2014){Lagos}, {Baugh}, {Zwaan}, {Lacey},
  {Gonzalez-Perez}, {Power}, {Swinbank}, \& {van Kampen}}]{Lagos:2014}
{Lagos} C.~D.~P., {Baugh} C.~M., {Zwaan} M.~A., {Lacey} C.~G., {Gonzalez-Perez}
  V., {Power} C., {Swinbank} A.~M., {van Kampen} E., 2014, MNRAS, 440, 920

\bibitem[{{Lah} {et~al}\mbox{.}(2009){Lah}, {Pracy}, {Chengalur}, {Briggs},
  {Colless}, {de Propris}, {Ferris}, {Schmidt}, \& {Tucker}}]{Lah:2009}
{Lah} P. {et~al.}, 2009, MNRAS, 399, 1447

\bibitem[{{Lamareille}(2010)}]{Lamareille:2010}
{Lamareille} F., 2010, A\&A, 509, A53

\bibitem[{{Large} {et~al}\mbox{.}(1981){Large}, {Mills}, {Little}, {Crawford},
  \& {Sutton}}]{Large:1981}
{Large} M.~I., {Mills} B.~Y., {Little} A.~G., {Crawford} D.~F., {Sutton} J.~M.,
  1981, MNRAS, 194, 693

\bibitem[{{Lewis} \& {Ibata}(2003)}]{Lewis:2003}
{Lewis} G.~F., {Ibata} R.~A., 2003, MNRAS, 340, 562

\bibitem[{{Liu} {et~al}\mbox{.}(2013){Liu}, {Zakamska}, {Greene}, {Nesvadba},
  \& {Liu}}]{Liu:2013}
{Liu} G., {Zakamska} N.~L., {Greene} J.~E., {Nesvadba} N.~P.~H., {Liu} X.,
  2013, MNRAS, 436, 2576

\bibitem[{{Liu} {et~al}\mbox{.}(2010){Liu}, {Shen}, {Strauss}, \&
  {Greene}}]{Liu:2010}
{Liu} X., {Shen} Y., {Strauss} M.~A., {Greene} J.~E., 2010, ApJ, 708, 427

\bibitem[{{L{\'o}pez-Caniego} {et~al}\mbox{.}(2007){L{\'o}pez-Caniego},
  {Gonz{\'a}lez-Nuevo}, {Herranz}, {Massardi}, {Sanz}, {De Zotti},
  {Toffolatti}, \& {Arg{\"u}eso}}]{Lopez-Caniego:2007}
{L{\'o}pez-Caniego} M., {Gonz{\'a}lez-Nuevo} J., {Herranz} D., {Massardi} M.,
  {Sanz} J.~L., {De Zotti} G., {Toffolatti} L., {Arg{\"u}eso} F., 2007, ApJS,
  170, 108

\bibitem[{{Macquart}(2005)}]{Macquart:2005}
{Macquart} J.-P., 2005, A\&A, 433, 827

\bibitem[{{Mahony} {et~al}\mbox{.}(2013){Mahony}, {Morganti}, {Emonts},
  {Oosterloo}, \& {Tadhunter}}]{Mahony:2013}
{Mahony} E.~K., {Morganti} R., {Emonts} B.~H.~C., {Oosterloo} T.~A.,
  {Tadhunter} C., 2013, MNRAS, 435, L58

\bibitem[{{Massardi} {et~al}\mbox{.}(2008){Massardi}, {Ekers}, {Murphy},
  {Ricci}, {Sadler}, {Burke}, {de Zotti}, {Edwards}, {Hancock}, {Jackson},
  {Kesteven}, {Mahony}, {Phillips}, {Staveley-Smith}, {Subrahmanyan}, {Walker},
  \& {Wilson}}]{Massardi:2008}
{Massardi} M. {et~al.}, 2008, MNRAS, 384, 775

\bibitem[{{Masui} {et~al}\mbox{.}(2013){Masui}, {Switzer}, {Banavar},
  {Bandura}, {Blake}, {Calin}, {Chang}, {Chen}, {Li}, {Liao}, {Natarajan},
  {Pen}, {Peterson}, {Shaw}, \& {Voytek}}]{Masui:2013}
{Masui} K.~W. {et~al.}, 2013, ApJ, 763, L20

\bibitem[{{Mauch} {et~al}\mbox{.}(2003){Mauch}, {Murphy}, {Buttery}, {Curran},
  {Hunstead}, {Piestrzynski}, {Robertson}, \& {Sadler}}]{Mauch:2003}
{Mauch} T., {Murphy} T., {Buttery} H.~J., {Curran} J., {Hunstead} R.~W.,
  {Piestrzynski} B., {Robertson} J.~G., {Sadler} E.~M., 2003, MNRAS, 342, 1117

\bibitem[{{McConnell} {et~al}\mbox{.}(2012){McConnell}, {Sadler}, {Murphy}, \&
  {Ekers}}]{McConnell:2012}
{McConnell} D., {Sadler} E.~M., {Murphy} T., {Ekers} R.~D., 2012, MNRAS, 422,
  1527

\bibitem[{{McElroy} {et~al}\mbox{.}(2015){McElroy}, {Croom}, {Pracy}, {Sharp},
  {Ho}, \& {Medling}}]{McElroy:2015}
{McElroy} R., {Croom} S.~M., {Pracy} M., {Sharp} R., {Ho} I.-T., {Medling}
  A.~M., 2015, MNRAS, 446, 2186

\bibitem[{{McMullin} {et~al}\mbox{.}(2007){McMullin}, {Waters}, {Schiebel},
  {Young}, \& {Golap}}]{McMullin:2007}
{McMullin} J.~P., {Waters} B., {Schiebel} D., {Young} W., {Golap} K., 2007, in
  ASP Conf. Ser., Vol. 376, Astronomical Data Analysis Software and Systems
  XVI, {Shaw} R.~A., {Hill} F., {Bell} D.~J., eds., Astron. Soc. Pac., San
  Francisco, p. 127

\bibitem[{{Meyer} {et~al}\mbox{.}(2004){Meyer}, {Zwaan}, {Webster},
  {Staveley-Smith}, {Ryan-Weber}, {Drinkwater}, {Barnes}, {Howlett}, {Kilborn},
  {Stevens}, {Waugh}, {Pierce}, {Bhathal}, {de Blok}, {Disney}, {Ekers},
  {Freeman}, {Garcia}, {Gibson}, {Harnett}, {Henning}, {Jerjen}, {Kesteven},
  {Knezek}, {Koribalski}, {Mader}, {Marquarding}, {Minchin}, {O'Brien},
  {Oosterloo}, {Price}, {Putman}, {Ryder}, {Sadler}, {Stewart}, {Stootman}, \&
  {Wright}}]{Meyer:2004}
{Meyer} M.~J. {et~al.}, 2004, MNRAS, 350, 1195

\bibitem[{{Moffet}(1975)}]{Moffet:1975}
{Moffet} A.~T., 1975, in Stars and Stellar Systems, Vol.~9, Galaxies and the
  Universe, {Sandage} A., {Sandage} M., {Kristian} J., eds., University of
  Chicago Press, Chicago, p. 211

\bibitem[{{Morganti} {et~al}\mbox{.}(2013){Morganti}, {Fogasy}, {Paragi},
  {Oosterloo}, \& {Orienti}}]{Morganti:2013}
{Morganti} R., {Fogasy} J., {Paragi} Z., {Oosterloo} T., {Orienti} M., 2013,
  {Science}, 341, 1082

\bibitem[{{Morganti} {et~al}\mbox{.}(2011){Morganti}, {Holt}, {Tadhunter},
  {Ramos Almeida}, {Dicken}, {Inskip}, {Oosterloo}, \&
  {Tzioumis}}]{Morganti:2011}
{Morganti} R., {Holt} J., {Tadhunter} C., {Ramos Almeida} C., {Dicken} D.,
  {Inskip} K., {Oosterloo} T., {Tzioumis} T., 2011, A\&A, 535, A97

\bibitem[{{Morganti} {et~al}\mbox{.}(2001){Morganti}, {Oosterloo}, {Tadhunter},
  {van Moorsel}, {Killeen}, \& {Wills}}]{Morganti:2001}
{Morganti} R., {Oosterloo} T.~A., {Tadhunter} C.~N., {van Moorsel} G.,
  {Killeen} N., {Wills} K.~A., 2001, MNRAS, 323, 331

\bibitem[{{Morganti} {et~al}\mbox{.}(2005){Morganti}, {Tadhunter}, \&
  {Oosterloo}}]{Morganti:2005b}
{Morganti} R., {Tadhunter} C.~N., {Oosterloo} T.~A., 2005, A\&A, 444, L9

\bibitem[{{Murgia}(2003)}]{Murgia:2003}
{Murgia} M., 2003, Publ. Astron. Soc. Aust., 20, 19

\bibitem[{{Murphy} {et~al}\mbox{.}(2007){Murphy}, {Mauch}, {Green}, {Hunstead},
  {Piestrzynska}, {Kels}, \& {Sztajer}}]{Murphy:2007}
{Murphy} T., {Mauch} T., {Green} A., {Hunstead} R.~W., {Piestrzynska} B.,
  {Kels} A.~P., {Sztajer} P., 2007, MNRAS, 382, 382

\bibitem[{{Noterdaeme} {et~al}\mbox{.}(2012){Noterdaeme}, {Petitjean},
  {Carithers}, {P{\^a}ris}, {Font-Ribera}, {Bailey}, {Aubourg}, {Bizyaev},
  {Ebelke}, {Finley}, {Ge}, {Malanushenko}, {Malanushenko},
  {Miralda-Escud{\'e}}, {Myers}, {Oravetz}, {Pan}, {Pieri}, {Ross},
  {Schneider}, {Simmons}, \& {York}}]{Noterdaeme:2012}
{Noterdaeme} P. {et~al.}, 2012, A\&A, 547, L1

\bibitem[{{Noterdaeme} {et~al}\mbox{.}(2009){Noterdaeme}, {Petitjean},
  {Ledoux}, \& {Srianand}}]{Noterdaeme:2009}
{Noterdaeme} P., {Petitjean} P., {Ledoux} C., {Srianand} R., 2009, A\&A, 505,
  1087

\bibitem[{{O'Dea}(1998)}]{Odea:1998}
{O'Dea} C.~P., 1998, PASP, 110, 493

\bibitem[{{Offringa} {et~al}\mbox{.}(2010){Offringa}, {de Bruyn}, {Biehl},
  {Zaroubi}, {Bernardi}, \& {Pandey}}]{Offringa:2010}
{Offringa} A.~R., {de Bruyn} A.~G., {Biehl} M., {Zaroubi} S., {Bernardi} G.,
  {Pandey} V.~N., 2010, MNRAS, 405, 155

\bibitem[{{Oosterloo} {et~al}\mbox{.}(2010){Oosterloo}, {Morganti}, {Crocker},
  {J{\"u}tte}, {Cappellari}, {de Zeeuw}, {Krajnovi{\'c}}, {McDermid},
  {Kuntschner}, {Sarzi}, \& {Weijmans}}]{Oosterloo:2010b}
{Oosterloo} T. {et~al.}, 2010, MNRAS, 409, 500

\bibitem[{{Oosterloo} {et~al}\mbox{.}(2009){Oosterloo}, {Verheijen}, {van
  Cappellen}, {Bakker}, {Heald}, \& {Ivashina}}]{Oosterloo:2009}
{Oosterloo} T., {Verheijen} M.~A.~W., {van Cappellen} W., {Bakker} L., {Heald}
  G., {Ivashina} M., 2009, in Wide Field Astronomy and Technology for the
  Square Kilometre Array, {Torchinsky} S.~A., {van Ardenne} A., {van den
  Brink-Havinga} T., {van Es} A.~J.~J., {Faulkner} A.~J., eds., Proc. Sci.

\bibitem[{{Orienti} {et~al}\mbox{.}(2006){Orienti}, {Morganti}, \&
  {Dallacasa}}]{Orienti:2006}
{Orienti} M., {Morganti} R., {Dallacasa} D., 2006, A\&A, 457, 531

\bibitem[{{Owsianik} \& {Conway}(1998)}]{Owsianik:1998}
{Owsianik} I., {Conway} J.~E., 1998, A\&A, 337, 69

\bibitem[{{P{\'e}roux} {et~al}\mbox{.}(2003){P{\'e}roux}, {McMahon},
  {Storrie-Lombardi}, \& {Irwin}}]{Peroux:2003}
{P{\'e}roux} C., {McMahon} R.~G., {Storrie-Lombardi} L.~J., {Irwin} M.~J.,
  2003, MNRAS, 346, 1103

\bibitem[{{Petrov}(2013)}]{Petrov:2013}
{Petrov} L., 2013, AJ, 146, 5

\bibitem[{{Pihlstr{\"o}m} {et~al}\mbox{.}(2003){Pihlstr{\"o}m}, {Conway}, \&
  {Vermeulen}}]{Pihlstrom:2003}
{Pihlstr{\"o}m} Y.~M., {Conway} J.~E., {Vermeulen} R.~C., 2003, A\&A, 404, 871

\bibitem[{{Planck Collaboration VII}(2011)}]{Planck:2011}
{Planck Collaboration VII}, 2011, A\&A, 536, A7

\bibitem[{{Randall} {et~al}\mbox{.}(2011){Randall}, {Hopkins}, {Norris}, \&
  {Edwards}}]{Randall:2011}
{Randall} K.~E., {Hopkins} A.~M., {Norris} R.~P., {Edwards} P.~G., 2011, MNRAS,
  416, 1135

\bibitem[{{Rao} {et~al}\mbox{.}(2006){Rao}, {Turnshek}, \& {Nestor}}]{Rao:2006}
{Rao} S.~M., {Turnshek} D.~A., {Nestor} D.~B., 2006, ApJ, 636, 610

\bibitem[{{Readhead} {et~al}\mbox{.}(1996){Readhead}, {Taylor}, {Xu},
  {Pearson}, {Wilkinson}, \& {Polatidis}}]{Readhead:1996}
{Readhead} A.~C.~S., {Taylor} G.~B., {Xu} W., {Pearson} T.~J., {Wilkinson}
  P.~N., {Polatidis} A.~G., 1996, ApJ, 460, 612

\bibitem[{{Reynolds}(1994)}]{Reynolds:1994}
{Reynolds} J., 1994, AT Technical Document AT/39.3/040

\bibitem[{{Rhee} {et~al}\mbox{.}(2013){Rhee}, {Zwaan}, {Briggs}, {Chengalur},
  {Lah}, {Oosterloo}, \& {Hulst}}]{Rhee:2013}
{Rhee} J., {Zwaan} M.~A., {Briggs} F.~H., {Chengalur} J.~N., {Lah} P.,
  {Oosterloo} T., {Hulst} T.~v.~d., 2013, MNRAS, 435, 2693

\bibitem[{{Ricci} {et~al}\mbox{.}(2006){Ricci}, {Prandoni}, {Gruppioni},
  {Sault}, \& {de Zotti}}]{Ricci:2006}
{Ricci} R., {Prandoni} I., {Gruppioni} C., {Sault} R.~J., {de Zotti} G., 2006,
  A\&A, 445, 465

\bibitem[{{Rosario} {et~al}\mbox{.}(2010){Rosario}, {Shields}, {Taylor},
  {Salviander}, \& {Smith}}]{Rosario:2010}
{Rosario} D.~J., {Shields} G.~A., {Taylor} G.~B., {Salviander} S., {Smith}
  K.~L., 2010, ApJ, 716, 131

\bibitem[{{Sault} {et~al}\mbox{.}(1995){Sault}, {Teuben}, \&
  {Wright}}]{Sault:1995}
{Sault} R.~J., {Teuben} P.~J., {Wright} M.~C.~H., 1995, in ASP Conf. Ser.,
  Vol.~77, Astronomical Data Analysis Software and Systems IV, {Shaw} R.~A.,
  {Payne} H.~E., {Hayes} J.~J.~E., eds., Astron. Soc. Pac., San Francisco, p.
  433

\bibitem[{{Schinckel} {et~al}\mbox{.}(2012){Schinckel}, {Bunton}, {Cornwell},
  {Feain}, \& {Hay}}]{Schinckel:2012}
{Schinckel} A.~E., {Bunton} J.~D., {Cornwell} T.~J., {Feain} I., {Hay} S.~G.,
  2012, in Proc. SPIE, Vol. 8444, 84442A

\bibitem[{{Serra} {et~al}\mbox{.}(2012){Serra}, {Oosterloo}, {Morganti},
  {Alatalo}, {Blitz}, {Bois}, {Bournaud}, {Bureau}, {Cappellari}, {Crocker},
  {Davies}, {Davis}, {de Zeeuw}, {Duc}, {Emsellem}, {Khochfar},
  {Krajnovi{\'c}}, {Kuntschner}, {Lablanche}, {McDermid}, {Naab}, {Sarzi},
  {Scott}, {Trager}, {Weijmans}, \& {Young}}]{Serra:2012}
{Serra} P. {et~al.}, 2012, MNRAS, 422, 1835

\bibitem[{{Shen} {et~al}\mbox{.}(2011){Shen}, {Liu}, {Greene}, \&
  {Strauss}}]{Shen:2011}
{Shen} Y., {Liu} X., {Greene} J.~E., {Strauss} M.~A., 2011, ApJ, 735, 48

\bibitem[{{Smith} {et~al}\mbox{.}(2010){Smith}, {Shields}, {Bonning},
  {McMullen}, {Rosario}, \& {Salviander}}]{Smith:2010}
{Smith} K.~L., {Shields} G.~A., {Bonning} E.~W., {McMullen} C.~C., {Rosario}
  D.~J., {Salviander} S., 2010, ApJ, 716, 866

\bibitem[{{Snellen} {et~al}\mbox{.}(2000){Snellen}, {Schilizzi}, \& {van
  Langevelde}}]{Snellen:2000}
{Snellen} I.~A.~G., {Schilizzi} R.~T., {van Langevelde} H.~J., 2000, MNRAS,
  319, 429

\bibitem[{{Stanghellini} {et~al}\mbox{.}(1997){Stanghellini}, {O'Dea}, {Baum},
  {Dallacasa}, {Fanti}, \& {Fanti}}]{Stanghellini:1997}
{Stanghellini} C., {O'Dea} C.~P., {Baum} S.~A., {Dallacasa} D., {Fanti} R.,
  {Fanti} C., 1997, A\&A, 325, 943

\bibitem[{{Stockton} {et~al}\mbox{.}(2007){Stockton}, {Canalizo}, {Fu}, \&
  {Keel}}]{Stockton:2007}
{Stockton} A., {Canalizo} G., {Fu} H., {Keel} W., 2007, ApJ, 659, 195

\bibitem[{{Tadhunter} {et~al}\mbox{.}(2014){Tadhunter}, {Morganti}, {Rose},
  {Oonk}, \& {Oosterloo}}]{Tadhunter:2014}
{Tadhunter} C., {Morganti} R., {Rose} M., {Oonk} J.~B.~R., {Oosterloo} T.,
  2014, {Nature}, 511, 440

\bibitem[{{Tengstrand} {et~al}\mbox{.}(2009){Tengstrand}, {Guainazzi},
  {Siemiginowska}, {Fonseca Bonilla}, {Labiano}, {Worrall}, {Grandi}, \&
  {Piconcelli}}]{Tengstrand:2009}
{Tengstrand} O., {Guainazzi} M., {Siemiginowska} A., {Fonseca Bonilla} N.,
  {Labiano} A., {Worrall} D.~M., {Grandi} P., {Piconcelli} E., 2009, A\&A, 501,
  89

\bibitem[{{Tingay} {et~al}\mbox{.}(2003){Tingay}, {Jauncey}, {King},
  {Tzioumis}, {Lovell}, \& {Edwards}}]{Tingay:2003}
{Tingay} S.~J., {Jauncey} D.~L., {King} E.~A., {Tzioumis} A.~K., {Lovell}
  J.~E.~J., {Edwards} P.~G., 2003, PASJ, 55, 351

\bibitem[{{van der Hulst} {et~al}\mbox{.}(2001){van der Hulst}, {van Albada},
  \& {Sancisi}}]{VanDerHulst:2001}
{van der Hulst} J.~M., {van Albada} T.~S., {Sancisi} R., 2001, in ASP Conf.
  Ser., Vol. 240, Gas and Galaxy Evolution, {Hibbard} J.~E., {Rupen} M., {van
  Gorkom} J.~H., eds., Astron. Soc. Pac., San Francisco, p. 451

\bibitem[{{van Gorkom} {et~al}\mbox{.}(1989){van Gorkom}, {Knapp}, {Ekers},
  {Ekers}, {Laing}, \& {Polk}}]{VanGorkom:1989}
{van Gorkom} J.~H., {Knapp} G.~R., {Ekers} R.~D., {Ekers} D.~D., {Laing} R.~A.,
  {Polk} K.~S., 1989, AJ, 97, 708

\bibitem[{{Vazdekis} {et~al}\mbox{.}(2010){Vazdekis},
  {S{\'a}nchez-Bl{\'a}zquez}, {Falc{\'o}n-Barroso}, {Cenarro}, {Beasley},
  {Cardiel}, {Gorgas}, \& {Peletier}}]{Vazdekis:2010}
{Vazdekis} A., {S{\'a}nchez-Bl{\'a}zquez} P., {Falc{\'o}n-Barroso} J.,
  {Cenarro} A.~J., {Beasley} M.~A., {Cardiel} N., {Gorgas} J., {Peletier}
  R.~F., 2010, MNRAS, 404, 1639

\bibitem[{{Verheijen} {et~al}\mbox{.}(2010){Verheijen}, {Deshev}, {van Gorkom},
  {Poggianti}, {Chung}, {Cybulski}, {Dwarakanath}, {Montero-Casta{\~n}o},
  {Morrison}, {Schiminovich}, {Szomoru}, \& {Yun}}]{Verheijen:2010}
{Verheijen} M.~A.~W. {et~al.}, 2010, in ISKAF2010 Science Meeting, {van
  Leeuwen} J., ed., Proc. Sci.

\bibitem[{{Vermeulen} {et~al}\mbox{.}(2003){Vermeulen}, {Pihlstr{\"o}m},
  {Tschager}, {de Vries}, {Conway}, {Barthel}, {Baum}, {Braun}, {Bremer},
  {Miley}, {O'Dea}, {R{\"o}ttgering}, {Schilizzi}, {Snellen}, \&
  {Taylor}}]{Vermeulen:2003}
{Vermeulen} R.~C. {et~al.}, 2003, A\&A, 404, 861

\bibitem[{{Wall} \& {Peacock}(1985)}]{Wall:1985}
{Wall} J.~V., {Peacock} J.~A., 1985, MNRAS, 216, 173

\bibitem[{{Wall} {et~al}\mbox{.}(1975){Wall}, {Shimmins}, \&
  {Bolton}}]{Wall:1975}
{Wall} J.~V., {Shimmins} A.~J., {Bolton} J.~G., 1975, Aust. J. Phys. Astrophys.
  Suppl., 34, 55

\bibitem[{{Walter} {et~al}\mbox{.}(2008){Walter}, {Brinks}, {de Blok},
  {Bigiel}, {Kennicutt}, {Thornley}, \& {Leroy}}]{Walter:2008}
{Walter} F., {Brinks} E., {de Blok} W.~J.~G., {Bigiel} F., {Kennicutt}, Jr.
  R.~C., {Thornley} M.~D., {Leroy} A., 2008, AJ, 136, 2563

\bibitem[{{Wang} {et~al}\mbox{.}(2013){Wang}, {Kauffmann}, {J{\'o}zsa},
  {Serra}, {van der Hulst}, {Bigiel}, {Brinchmann}, {Verheijen}, {Oosterloo},
  {Wang}, {Li}, {den Heijer}, \& {Kerp}}]{Wang:2013}
{Wang} J. {et~al.}, 2013, MNRAS, 433, 270

\bibitem[{{Wolfe} {et~al}\mbox{.}(1982){Wolfe}, {Briggs}, \&
  {Davis}}]{Wolfe:1982}
{Wolfe} A.~M., {Briggs} F.~H., {Davis} M.~M., 1982, ApJ, 259, 495

\bibitem[{{Wolfire} {et~al}\mbox{.}(2003){Wolfire}, {McKee}, {Hollenbach}, \&
  {Tielens}}]{Wolfire:2003}
{Wolfire} M.~G., {McKee} C.~F., {Hollenbach} D., {Tielens} A.~G.~G.~M., 2003,
  ApJ, 587, 278

\bibitem[{{Wright} {et~al}\mbox{.}(1994){Wright}, {Griffith}, {Burke}, \&
  {Ekers}}]{Wright:1994}
{Wright} A.~E., {Griffith} M.~R., {Burke} B.~F., {Ekers} R.~D., 1994, ApJS, 91,
  111

\bibitem[{{Wright} {et~al}\mbox{.}(2010){Wright}, {Eisenhardt}, {Mainzer},
  {Ressler}, {Cutri}, {Jarrett}, {Kirkpatrick}, {Padgett}, {McMillan},
  {Skrutskie}, {Stanford}, {Cohen}, {Walker}, {Mather}, {Leisawitz}, {Gautier},
  {McLean}, {Benford}, {Lonsdale}, {Blain}, {Mendez}, {Irace}, {Duval}, {Liu},
  {Royer}, {Heinrichsen}, {Howard}, {Shannon}, {Kendall}, {Walsh}, {Larsen},
  {Cardon}, {Schick}, {Schwalm}, {Abid}, {Fabinsky}, {Naes}, \&
  {Tsai}}]{Wright:2010}
{Wright} E.~L. {et~al.}, 2010, AJ, 140, 1868

\bibitem[{{Zafar} {et~al}\mbox{.}(2013){Zafar}, {P{\'e}roux}, {Popping},
  {Milliard}, {Deharveng}, \& {Frank}}]{Zafar:2013}
{Zafar} T., {P{\'e}roux} C., {Popping} A., {Milliard} B., {Deharveng} J.-M.,
  {Frank} S., 2013, A\&A, 556, A141

\bibitem[{{Zwaan} {et~al}\mbox{.}(2005){Zwaan}, {van der Hulst}, {Briggs},
  {Verheijen}, \& {Ryan-Weber}}]{Zwaan:2005}
{Zwaan} M.~A., {van der Hulst} J.~M., {Briggs} F.~H., {Verheijen} M.~A.~W.,
  {Ryan-Weber} E.~V., 2005, MNRAS, 364, 1467

\end{thebibliography}

\section*{Author Affiliations}
$^{1}$CSIRO Astronomy and Space Science, PO Box 76, Epping, NSW 1710,
Australia\\$^{2}$Sydney Institute for Astronomy, School of Physics
A28, University of Sydney, Sydney, NSW 2006, Australia\\$^{3}$ARC
Centre of Excellence for All-sky Astrophysics (CAASTRO)\\$^{4}$School
of Chemical and Physical Sciences, Victoria University of Wellington,
PO Box 600, Wellington 6140, New Zealand\\$^{5}$Netherlands Institute
for Radio Astronomy, Postbus 2, NL-7990 AA Dwingeloo, the
Netherlands\\$^{6}$Kapteyn Astronomical Institute, University of
Groningen, Postbus 800, NL-9700 AV Groningen, the
Netherlands\\$^{7}$School of Physical Sciences, University of
Tasmania, Private Bag 37, Hobart Tasmania 7001,
Australia\\$^{8}$European Southern Observatory,
Karl-Schwarzschild-Str. 2, D-85748 Garching, Germany\\$^{9}$SKA
Organisation, Jodrell Bank Observatory, Lower Withington,
Macclesfield, Cheshire, SK11 9DL, UK\\$^{10}$CSIRO Digital
Productivity, PO Box 76, Epping, NSW 1710, Australia\\$^{11}$Radio
Astronomy Laboratory, University of California Berkeley, 501 Campbell,
Berkeley CA 94720-3411, USA\\$^{12}$Radiation Physics Laboratory,
Sydney Medical School, The University of Sydney, NSW 2006,
Australia\\$^{13}$Inter-University Centre for Astronomy and
Astrophysics, Post Bag 4, Ganeshkhind, Pune University Campus, Pune
411 007, India\\$^{14}$Department of Physics and Electronics, Rhodes
University, PO Box 94, Grahamstown 6140, South
Africa\\$^{15}$International Centre for Radio Astronomy Research
(ICRAR), Curtin University, GPO Box U1987, Perth, WA 6845,
Australia\\$^{16}$Research School of Astronomy and Astrophysics,
Australian National University, Mount Stromlo Observatory, Cotter
Road, Weston Creek, ACT 2611, Australia\\$^{17}$International Centre
for Radio Astronomy Research (ICRAR), The University of Western
Australia, 35 Stirling Hwy, Crawley, WA 6009, Australia\\$^{18}$School
of Physics, University of Melbourne, Victoria 3010,
Australia\\$^{19}$Leiden Observatory, Leiden University, PO Box 9513,
NL-2300 RA Leiden, the Netherlands

\label{lastpage}

\end{document}